**Policy Convergence and Divergence Across National and Within Regional AI Strategies: A Policy Design Element Analysis**

Benjamin Faveri, CEIMIA and Carleton University

Brie Bhasin, University of Ottawa

## Abstract

Governments worldwide have responded to the rapid expansion of AI by publishing national and regional AI strategies. Comparing national and regional AI strategies to identify their convergences and divergences can uncover their common practices, understand regional variations, and provide policy designers a comprehensive set of policy design elements for their ongoing AI strategy developments. Yet, existing work has not examined their underlying policy design elements or assessed whether those elements are horizontally (country-to-country) or vertical (region-to-country) converging or diverging over time. This paper addresses that gap by coding and analyzing 74 national and 3 regional AI strategies drawn from a global scan of all 205 UN member and non-member states. The coding used a latent-inductive approach organized around three functional policy design elements: goals, approaches, and principles. Two research questions guided the analysis: to what degree are national AI strategies becoming horizontally convergent or divergent over time; and to what degree are national strategies becoming vertically convergent or divergent with those countries' regional AI strategy. Results indicate strong horizontal convergence around economic competitiveness, research support, and ethical AI use, alongside persistent divergence in human rights goals, participatory governance approaches, and human-centric principles. Across the three regions, the AU exhibits the highest vertical convergence, the EU demonstrated strong alignment on regulatory and economic priorities but diverges on human-centric values, and the Nordic-Baltic Region displays mixed vertical convergence. These findings offer policy designers a comprehensive evidence base for identifying emerging AI policy design choice norms as AI strategies are developed and updated.

## Introduction

Artificial intelligence (AI) has emerged as one of the most consequential policy challenges in contemporary public policy. Governments worldwide have recognized that the rapid expansion of AI across domains like healthcare, finance, labour markets, and national security, among many others, demands coordinated state-level responses. This recognition is reflected in a global trend of national AI strategies. Beginning in 2017, when Canada launched what is widely regarded as the first national AI strategy, a wave of national and regional AI strategies has followed, driven by growing recognition of AI's potential for national productivity, competitive advantage, and geopolitical influence. By the end of 2022, 68 countries had released national AI strategies, with the dataset assembled for this study ultimately identifying 95 published national AI strategies across all 205 UN member and non-member states, supplemented by three regional strategies from the European Union (EU), African Union (AU), and Nordic-Baltic Region (NBR) by the end-of-2025.

One might expect this to be a story of convergence. A novel technology emerges that demands governmental attention, and states respond with strategies that speak to similar themes regarding the economic, security, and citizen welfare implications and opportunities arising from what is now widely understood as a highly disruptive technology. Yet, divergence may be equally plausible and likely given the sheer scale and pace of AI development alongside competing national interests and varying institutional capacities that could leave governments choosing different priorities when strategizing about the way forward. Understanding whether and how countries are converging in their national and regional AI policy design efforts, and what that convergence or divergence means for the firms, governments, and people operating within those environments has become an increasingly pressing concern for policy designers. Yet, existing comparative research has largely documented common themes at a surface level without systematically disaggregating strategies according to their underlying policy design elements or assessing whether those elements are converging or diverging over time – a gap with consequences for policy designers (Dua et al., 2025; Filgueiras, 2022; Hjaltalin and Sigurdarson, 2024; Radu, 2021; Salas-Pilco, 2021; Singh et al., 2025; van Noordt et al., 2023).

This article attends to this issue by asking *how and to what extent national and regional AI strategies are converging or diverging?* The study adopts a policy design approach – where policy design refers to the deliberate selection and combination of elements within a policy instrument, particularly its -goals, -approaches, and -principles. Policy-goals define what a policy aims to accomplish; policy-approaches specify the mechanisms through which those goals are to be pursued; and policy-principles articulate the normative values and standards that should be upheld throughout the goal-attainment process. A well-designed policy typically integrates all three element types in a coherent and mutually reinforcing mix, though designers must remain attentive to redundant, contradictory, or counter-productive combinations that can undermine a policy's capacity to achieve its stated objectives (Howlett, 2014). Policy convergence is defined as the tendency for policies across distinct jurisdictions to be or become more alike over time in their structures and substantive commitments, a process that can occur horizontally (across jurisdictions at the same administrative level), or vertically (between different administrative levels within the same hierarchy, such as a regional body and its Member states). Importantly, convergence and divergence can coexist within the same domain, producing divergence-within-convergence, where jurisdictions adopt similar high-level governance language while diverging on the specific approaches and principles used to operationalize it. In this respect, the study will investigate convergence and divergence among national strategies and between national strategies and the regional strategies to which the country is allied.

The study's focus is motivated by a practical imperative as nations and regions continue to publish, update, and implement AI strategies, policy designers need an empirically grounded understanding of what design elements these strategies contain, how those elements relate across jurisdictions, and what trajectories of convergence or divergence are emerging as the field matures. To address these questions, a dataset of 74 national and 3 regional AI strategies was constructed through a systematic global scan, with each strategy coded using a latent-inductive approach based on their policy-goals, -approaches, and -principles at broad and specific levels, and the temporal evolution of codes charted to evaluate convergence and divergence patterns.

The results reveal a landscape characterized by strong convergence around certain economic and institutional priorities alongside persistent divergence in substantive social commitments. Across national strategies, near-complete convergence is observed around supporting national AI research (71/74 strategies) and ethical AI use (68/74 strategies), suggesting these elements have effectively become baseline expectations for any credible national AI strategy. At the same time, divergence persists around human-rights goals, participatory governance approaches, and human-centric principles like diversity and inclusion, pointing to a structural gap between commitments and the substantive choices that would operationalize those commitments. Within regions, the AU exhibits the highest vertical convergence, the EU demonstrates strong alignment on regulatory and economic goals but diverges on human-centric principles, and the NBR displays mixed vertical convergence with alignment around broad ethical values but divergence across operational goals and approaches. Together, these findings suggest that the window for independent AI policy problem-solving is narrowing, that ethics washing risks eroding public trust in strategies that adopt ethical language without substantive follow-through, and that regional strategies that are too abstract risk producing minimal national convergence, all of which has implications for how policy designers approach the development and updating of national and regional AI strategies.

The remainder of this paper develops in five stages. First, a three-part literature is provided in policy design, policy convergence and divergence, and the evolution of national and regional AI strategies from 2017 onward. Second, the methodology is detailed across how the dataset was created, how coding was conducted, and how the analysis was completed to answer the research questions. Third, results were detailed for each RQ. Fourth, a three-part discussion is offered around the policy design element variations, horizontal and vertical policy convergence and divergence implications, and the utility of this paper's findings for policy designers. Fifth, some concluding remarks are given around the overarching significance of this paper's findings, its limitations, and areas of future research.

## Responding to AI Through National and Regional Strategies

The proliferation of national and regional AI strategies represents a considerable development in contemporary public policy. Beginning with Canada's (2017) *Pan-Canadian AI Strategy*, widely regarded as the first national AI strategy, China and Finland quickly followed with their own national AI strategies in the same year. More national strategies followed, all of which, in some fashion, emerged in response to the growing recognition that AI's potential needed a coordinated state-level policy response, especially with its applications spanning healthcare diagnostics, labour market restructuring, public service delivery, and national security, among several other areas. Simultaneously, governments were aware of the competitive dynamics at stake, with AI increasingly framed as a pillar of national economic power and geopolitical influence (Auld et al., 2022; Mazarr, 2026). From 2017 to 2019, a foundational landscape was established in which a relatively small number of technologically advanced states defined the early contours of national AI policy, focusing their strategies on R&D investment, talent development, and the articulation of broad ethical commitments.

The pace of AI strategy publication increased considerably from 2019 onward, and particularly after 2020, as the broader AI policy community grappled with the implications of rapid machine learning developments and the emergence of generative AI. By the end-of-2020, 49 countries had released national AI strategies, and by the end-of-2022, that number had grown to 68. As documented in this study's dataset, 95 published national AI strategies were ultimately identified across all 205 UN member and non-member states, a figure supplemented by 30 strategies in progress and 5 called-for by their governments by the end-of-2025. Three regional AI strategies were also identified: the EU, AU, and NBR, representing distinct supranational or intergovernmental strategies designed to coordinate AI governance across member states. This expanding corpus of AI strategies suggests a broad consensus on the necessity of deliberate AI planning. In 2025, most major economies have either updated their earlier national AI strategies or are actively updating them, with second-generation plans giving greater weight to compute access, sovereign data, safety frameworks, and talent pipelines alongside the innovation and growth priorities that dominated earlier iterations, particularly as geopolitical and global economic tensions rise (see Appendix A for status of all national and regional AI strategies).

Despite this proliferation, the existing body of comparative research on national and regional AI strategies has remained somewhat limited in analytical depth. Prior work has documented the content of these strategies in summary form, identifying common calls for research investment, regulation, and data ethics, but this work has not systematically examined the variation in their underlying policy design elements or assessed whether these elements are converging or diverging over time. Fatima et al. (2020) analyzed 30+ national AI strategies to understand common themes and structural priorities. The OECD's (2026) *AI Policy Observatory* has catalogued over 1,000 policy initiatives across more than 70 jurisdictions. Roberts et al. (2023) produced comparative analyses of specific national cases, including China and the EU, highlighting differences in what these strategies aimed to accomplish, how AI development and use is promoted, and whom these policies are designed to benefit. The European Commission Joint Research Centre's (2020) report similarly produced a European perspective on national strategies benchmarked against the EU's (2018) *Coordinated Plan on AI* – which would later develop into what is now their (2025) regional AI strategy. What has been absent from this body of work is a systematic analysis that disaggregates these strategies according to a rigorous and replicable policy design framework and that maps whether their policy design elements are converging or diverging over time – a gap this paper directly addresses.

### Analytic Approach for Understanding Strategic Convergence and Divergence

The analytic approach adopted is informed by research on policy design and policy convergence and divergence. Policy design focuses on the different elements within a policy, namely, a policy's goals, objectives, aims, approaches, principles, calibrations, and settings (Howlett, 2014). These policy design elements are not always included in every policy; sometimes, policy design elements are missing from a policy. Missing some policy design elements does not necessarily make a policy worse than a policy with more elements. The policy design element's comprehensiveness will depend on the nature of the problem being addressed, the current government, policy designers, and several other factors (Howlett, 2019a). Typically, a policy will include a goal, one or more approaches to achieving that goal, and some underlying principles that must be upheld throughout the policy-goal attainment process. While Howlett (2019a) differentiates between policy goals, objectives, and aims, this paper groups these three terms together as "policy-goals" given their definitional overlap – all defining what a policy tries to accomplish. The slight definitional differences overcomplicate the general functional policy design theory expressed throughout this paper (see Figure 1 for functional policy design theory). Moreover, in the context of strategy documents, objectives and aims are less likely to be well represented, as they represent operational elements in programs of action (perhaps cite some work here that has applied the framework in other contexts where there are objectives and aims identified to show the concrete differences between them in those other contexts).

*Figure 1: Functional Policy Design Theory*

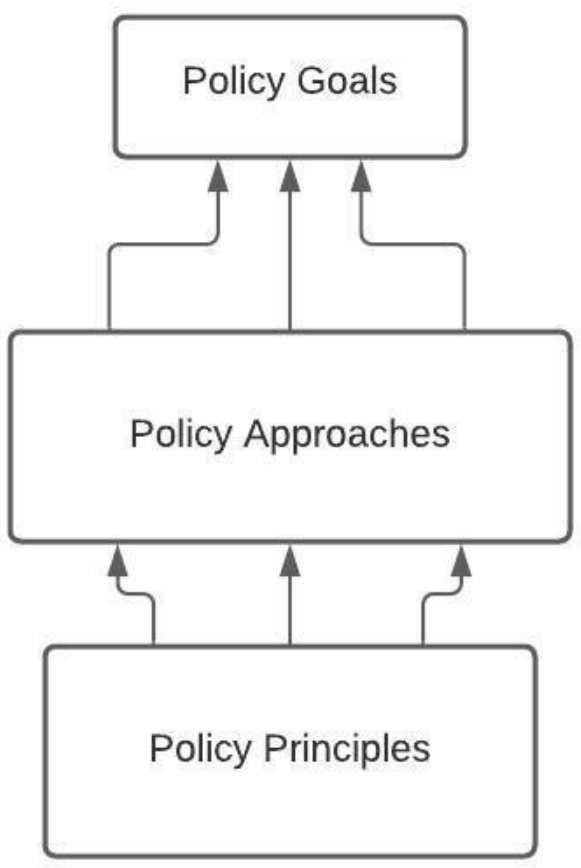

Howlett (2014; 2019a: 2019b) argues that policy designers must use several policy-approaches to achieve any given policy-goal. Each policy-approach used within a given policy will, in some way, rely on government spending, regulation, and political weight, among other resources, to achieve its policy-goal. Understanding the various policy design elements used to address a given policy problem can help future policy designers design more informed policies. Past studies have examined the advantages and disadvantages of various policy design elements around a given policy problem (Schneider & Sidney, 2009). However, choosing which policy-goals, -approaches, and -principles are "ideal" during the policy design process is often difficult as several policy designers, sectors, and other parties' interests' conflict as each party tries to influence the policy toward what that party thinks is ideal. With several parties trying to influence the policy design process, finding a mix of complementary policy design elements can often become the policy designer's central focus. Policy designers should try to avoid redundant, repetitive, and counter-productive policy design element mixes as these can hinder such policies' ability to attain their stated goal(s). Either way, policy designers must be aware of policy design element differences within a given policy area to effectively design their policies. One way for policy designers to understand and become aware of various policy design elements within a policy area is to compare other similar policies to get a sense of their policy convergence or divergence (Bennett, 1991; Drezner, 2001; Holzinger & Knill, 2005).

Policy convergence is usually defined as the tendency to *be* or *become more* alike in structures, processes, and performances, or other criteria (Bennett, 1991), and requires identifiable movement from different starting positions toward an identified common point over some period of time (Holzinger & Knill, 2005). Policy divergence is the reverse and tends to occur when political jurisdictions maintain differences or grow apart in their policy responses given their varying domestic institutions, political opportunity structures, and independent national traditions (Knill, 2005). Policies can also simultaneously exhibit a phenomenon of divergence within convergence where states may converge on adopting similar high-level governance efforts but diverge in substantive policy output production given distinct capacities of their domestic public and private actors (Howlett & Rayner, 2006).

A comprehensive understanding of policy convergence in a given policy area requires examining convergence and divergence across multiple administrative and spatial dimensions, specifically horizontal and vertical planes. Horizontal policy convergence refers to the growing similarity of policies across jurisdictions operating at the same administrative level (Zhu et al., 2025). Conversely, vertical policy convergence, evaluates the convergence between different administrative levels within the same hierarchy, such as federal-provincial-territorial or EU-level-to-member state relations. This horizontal or verticality can occur cross-nationally between states navigating similar policy areas, or sub-nationally between regional governments. For instance, local governments with comparable economies and technological resources may independently adopt similar implementation strategies despite existing in different national systems, highlighting horizontal convergence (Zhu et al., 2025). High vertical convergence can occur when local governments closely align their localized policy-goals, -approaches, and -principles with overarching national or supranational directives. Conversely, vertical divergence occurs when subnational entities exercise their autonomy, formulating independent initiatives that differ from central mandates to fit local contexts. Figure 2 below visually represents horizontal (country-to-country) and vertical (region-to-country(ies)) policy convergence.

*Figure 2: Vertical (regional to country) and horizontal (country to country) policy convergence*

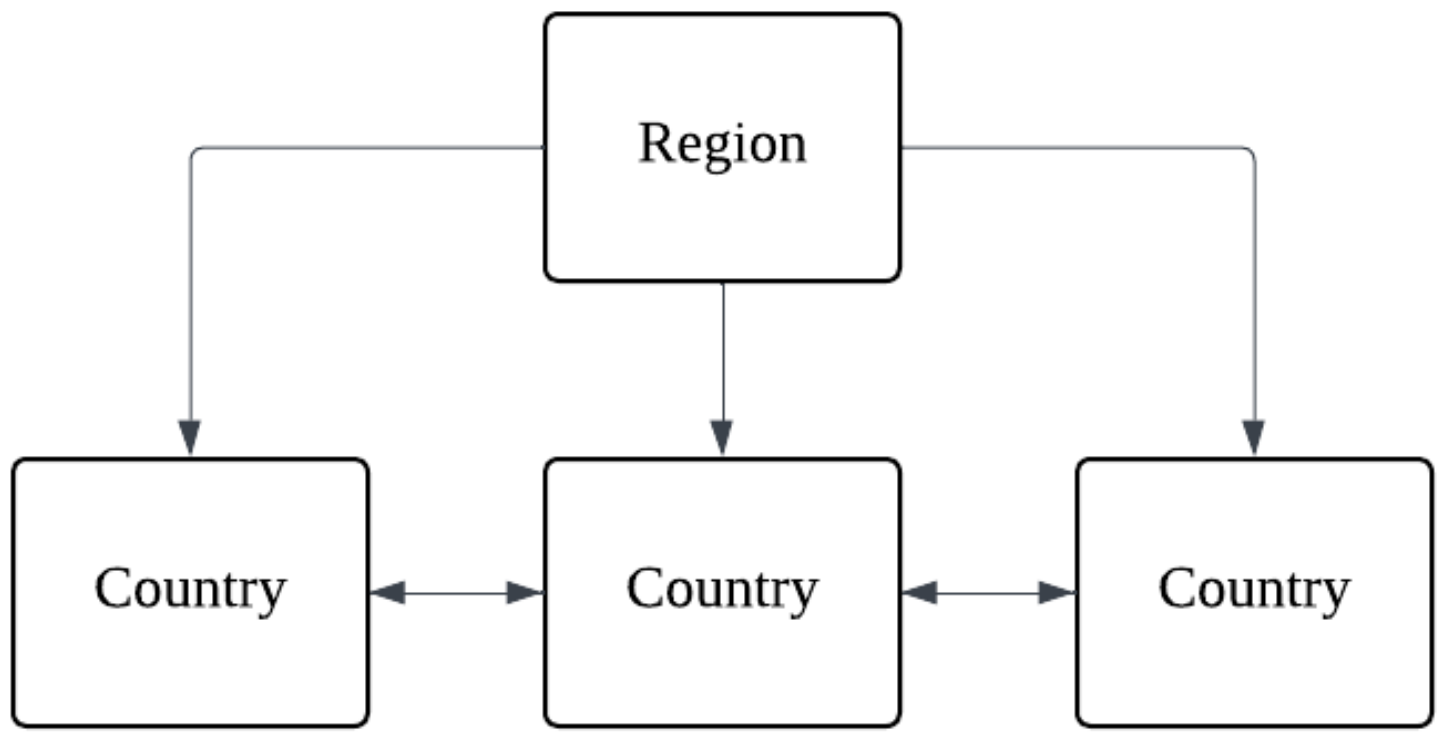

Measuring policy convergence is of critical importance as it can reveal early policy design element choices having a lasting effect on future policy design element choices; how governments are thinking about a given problem, if their thinking has changed, and when; and policy coalitions – whereby several jurisdictions take on the same or near the same policy design element choices that would compete or conflict with another policy coalition in the same policy area. (Drezner, 2001; Knill, 2005). By measuring convergence, analysts can examine such claims and determine whether similarities arise from external pressures, like harmonization, regulatory competition, or voluntary transnational communication, among other factors (Bennett, 1991; Holzinger & Knill, 2005). Understanding these dynamics is essential for policy designers, as it clarifies whether a given policy area is dominated by a few great powers establishing a global policy standard or if states are engaging in independent problem-solving (Drezner, 2005).

To accurately measure policy convergence and avoid the ambiguity of overly broad comparisons, researchers must systematically examine distinct policy design elements. Because public policy is a multidimensional phenomenon, convergence cannot be measured simply by looking at broad outcomes; it must be specified across several distinct categories, such as policy-goals, -approaches, and -principles (Howlett & Rayner, 2006). By qualitatively or quantitatively tracking these specific policy design elements, researchers can utilize precise metrics to evaluate policy convergence and divergence. For instance, analysts can measure the *degree* of convergence, often assessed as *s-convergence* (the decrease in standard deviation or variation of policies among countries over time); the *direction* of convergence (which tracks whether the regulatory mean shifts upward toward stricter standards or downward toward laxer rules); or the *scope* of convergence (which measures the total number of jurisdictions affected; Holzinger & Knill, 2005; Knill, 2005). Disaggregating policy into these measurable design elements allows researchers to identify nuanced realities. For example, Howlett & Rayner (2006) demonstrate that nations might converge on broad policy-goals and -principles, like sustainable environmental rhetoric, but diverge completely on the policy approaches and implementation instruments used to achieve them.

Finally, understanding the temporal evolution of policy convergence is paramount for policy designers as convergence is inherently a process of historical and sequential change. Without an explicit temporal framework, analysts risk inferring that a convergence or divergence is the result of a communication or coercion, when it could just be coincidental or an isolated occurrence of independent responses at the same time, or something else entirely (Bennett, 1991; Holzinger & Knill, 2005). Tracking temporal policy evolution allows policy designers to observe the cumulative adoption patterns. For policy designers, this temporal awareness is vital for policy adaptation as it allows them to engage in lesson-drawing from other jurisdictions' policy efforts – a process where they systematically analyze the experiences of first-mover states over time to design informed domestic frameworks (Bennett, 1991; Holzinger & Knill, 2005). Furthermore, mapping these trends temporally helps policy designers identify instances of policy layering, drift, or conversion – where old or contradictory policy instruments stack up over time – allowing them to proactively design cohesive and updated policies before one or more policy coalitions are formed and dominate the policy area (Drezner, 2005; Howlett & Rayner, 2006).

## Methodology

### *(a) Dataset Creation*

This dataset was created in three steps. First, a list of all 205 UN member and non-member countries was compiled. Second, each country on this list was searched to identify if it had, was currently developing, had called for, or did not have a national AI strategy (see Appendix A for a complete list of national AI strategies by their status).[1] When a country had a national AI strategy, the year it was published, the language it was written in, and a link to its published document (if possible) or article(s) supporting that is it in-development or called for were collected. During these searches, it was noticed that some countries had updated their national AI strategies. When these updates were identified, they were treated as additional national AI strategies. All strategy documents were downloaded and stored systematically, with national AI strategies explicitly marked to distinguish them from their original versions. To ensure consistent inclusion and exclusion, a clear definition of what constituted a "national AI strategy" was applied throughout these searches. A "national AI strategy" was defined as any document that was strictly about AI *and* broadly scoped to an entire country. This conceptualization excluded strategies that had AI as a part of them, such as digital strategies,[2] or strategies that were about a particular sector, industry, etc., like *Canada's AI Strategy for the*

[1] While there are existing compilations of national AI strategies, like the *Future of Life Institute* and *OECD.AI Policy Observatory*, they are either outdated or have a different conceptualization of "national AI strategy" than this paper. Granted this paper did search through these and other similar sources during steps 2 and 3.
[2] Example of a Digital Strategy: https://www.arab-digital-economy.org/2020/12.pdf

*Federal Public Service 2025-2027*[3] or the *Department of National Defence and Canadian Armed Forces AI Strategy*.[4] While many included national AI strategies were titled as "national AI strategies," other included strategies had titles like AI "action plans," "strategic programs", or "policy frameworks." These variously titled documents were included as they met both inclusion criteria. Third, during steps 1 and 2, three regions were identified to have "regional AI strategies." These regional AI strategies – the AU, EU and NBR – had the same information collected about them as the national AI strategy documents but were kept conceptually separate. Following the dataset's creation, 95 countries were identified as having a national AI strategy. From these 95 countries, countries without publicly accessible strategy documents and strategies written in non-English were excluded. When a country had an original and updated strategy(ies), all strategies were included in the final dataset. Following these exclusions and inclusions, 74 national and 3 regional AI strategies were included in the final dataset for coding and analysis (see Figure 3 for Sankey Diagram of national AI strategy results' inclusion and exclusion criteria).

*Figure 3: Sankey Diagram of National AI Strategy Inclusion and Exclusion Criteria Results*

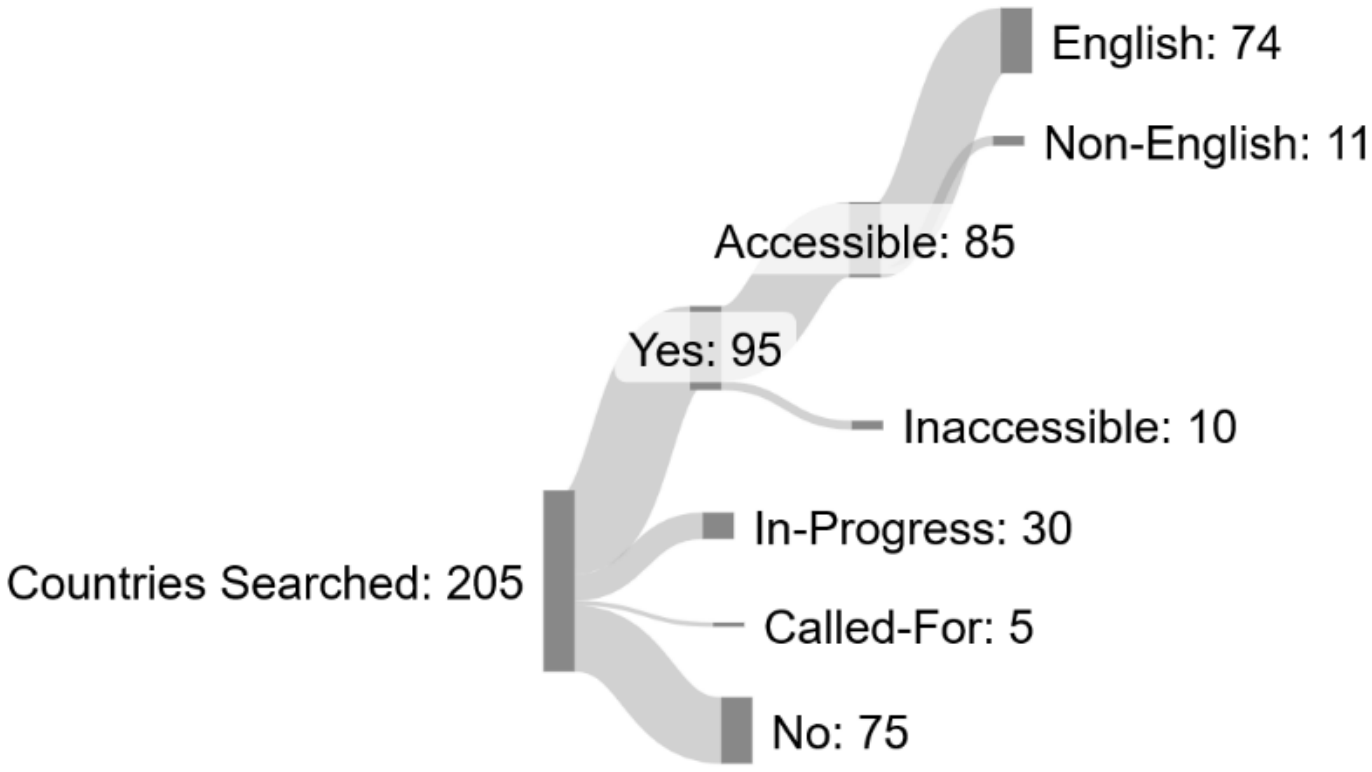

*(b) Coding Strategy*

To examine the policy-goals, -approaches, and -principles across national and regional AI strategies and allow for analysis of their relative horizontal and vertical policy convergence and divergence, this paper used a latent-inductive coding approach. This approach required reading each included strategy for their underlying broad and specific conceptual patterns within each policy design element. Broad codes refer to an overarching pattern while specific codes refer to patterns found within a given broad code pattern (e.g., "supporting national AI research" is a broad policy-approach code while "increasing the number of highly skilled AI workers and graduate students" is a specific code within that broad code). As these patterns were identified, they were added to the codebook under their relevant policy design element (see Appendix B for complete codebook). This board and specific coding categorization provided more detailed data for the subsequent analysis.

Once the codebook was completed, each of the strategies included were coded. This coding was completed by reading each strategy again to identify its relevant codes against the codebook. When a code was identified, the verbatim text supporting that code was lifted from the strategy document and placed into an Excel sheet with that text's corresponding code, allowing each coding decision to be supported by the strategy's text. For example, Spain's (2024) strategy[5] supported the broad code "Support National AI Research" and specific code "Increase High-Skilled AI Workers and Graduate Students" through its mention of "Fostering more public-private collaboration in research and the application of cutting-edge technologies" and "Several universities in Spain have created new courses (graduate and master's degrees) specializing in AI" and "Training grants for AI and digital enabling technologies." After all national and regional strategies were coded, the coding was analyzed to answer each research question.

[3] https://www.canada.ca/en/government/system/digital-government/digital-government-innovations/responsible-use-ai/gc-ai-strategy-overview.html
[4] https://publications.gc.ca/collections/collection_2024/mdn-dnd/D2-633-2024-eng.pdf

*(c) Analysis*

Once the coding was completed, the codes were prepared for analysis following two steps. First, all coded strategies were organized by their publication year, with updated strategies being included in subsequent years where applicable. Second, each coded strategy's verbatim text and corresponding broad and specific codes were translated into a 1 or 0, where 1 = code is present and 0 = code is absent from the strategy. Using these preparation steps, the data were analyzed to answer the research questions about: (1) horizontal divergence or convergence among national AI strategies and (2) vertical divergence or convergence between an individual national AI strategy and the regional AI strategy that the country aligned with geographically.

The analysis involved grouping AI strategies by publication year and summing specific codes for each policy design element. These totals were then graphed to illustrate shifts in policy design element usage overtime, indicating policy convergence or divergence when policy design element codes were becoming more or less frequent over time, when certain codes started to appear or disappear, and the relationship between a broad code and its specific codes usage shifted – which could indicate horizontal policy convergence or divergence within policy design elements. For the vertical patterns, applicable national AI strategies were grouped by their region (e.g., all EU-Member States with a national AI strategy were grouped together to analyze them against the EU's regional AI strategy). These strategies were grouped by publication year and had their codes totaled within each year. These totals were then graphically represented to illustrate shifts in vertical convergence relative to their regional AI strategy, following the same approach adopted for horizontal convergence.

**Results**

The analysis of 74 national and 3 regional AI strategies reveal a policy landscape characterized by simultaneous convergence and divergence across goals, approaches, and principles – a pattern of divergence-within-convergence that cuts across both horizontal and vertical dimensions. At the horizontal level, there is near-complete convergence observed around supporting national AI research (71/74 strategies) and ethical AI use (68/74 strategies), alongside strong and accelerating convergence around economic competitiveness and capacity-building goals, like becoming an AI leader (59/74) and building national capacity (61/74), suggesting these elements have effectively become baseline expectations for any credible national AI strategy. At the same time, persistent horizontal divergence appears in human rights goals (23/74), participatory governance approaches such as encouraging public consultations, regulatory efforts, and legislative frameworks, and human-centric principles including human dignity, diversity, and inclusion, which exposes a structural gap between the high-level ethical language adopted broadly and the substantive policy commitments that would operationalize such language. Turning to vertical patterns, the three regional strategies illustrate distinct profiles. The AU exhibits the highest vertical convergence across all three policy design elements, with member state strategies closely mirroring continental priorities around growth, societal benefit, ethical use, and data protection. The EU demonstrates strong alignment on regulatory frameworks and economic objectives but diverges on human-centric values, with the EC prioritizing diversity and inclusion principles that most member states underemphasize. And the NBR displays mixed vertical convergence, with shared broad ethical commitments but divergence across operational goals and approaches, suggesting its regional strategy functions more as a declaration of values than an operational governance framework.

---

[5]Spain's National AI Strategy: https://portal.mineco.gob.es/RecursosArticulo/mineco/ministerio/ficheros/National-Strategy-on-AI.pdf

Horizontal patterns show a general trend towards convergence of policy-goals over time (Figure 4). These data suggest a strong convergence around economic and capacity-building properties. “Build Capacity” (61/74 strategies) and “Become an AI Leader” (59/74 strategies) dominate the landscape, remaining consistently present across all years. This persistent emphasis indicates that countries are aligning with the view that national competitiveness and capacity-building are core AI policy-goals. For instance, Jamaica’s 2024 strategy stresses equipping citizens with digital literacy skills to ensure inclusive societal benefits, while Canada’s updated 2022 strategy highlights boosting productivity through research excellence. Both examples show how national strategies frame capacity-building as fundamental to long-term AI leadership, reinforcing horizontal convergence across contexts. Specific policy-goal codes also reveal areas of convergence. “Through Research and Development” appears in 56, while “Mentions Innovation” and “Through Skills” each appear in 49 strategies. These consistent inclusions point to a large consensus that investing in knowledge and skills is essential to remain competitive in AI. For example, Sweden’s 2024 strategy captures this consensus by linking AI adoption to societal resilience, while Italy’s 2024 strategy emphasizes leveraging AI for industrial competitiveness. Together, these examples illustrate how different countries converge on parallel priorities, further reinforcing convergence across national strategies.

However, this convergence is not universal. Human rights (23/74 strategies) and national security (15/74 strategies)[6] appear sporadically, suggesting divergence in how countries integrate broader social or security concerns into their AI policy-goals. For example, Zambia’s 2024 strategy highlights inclusivity and equitable benefits, but few other states mirror this emphasis. The limited uptake of these specific codes indicates divergence in socially oriented policy-goals, as most strategies prioritize economic and technological aims over rights-based or security-focused objectives, granted this may change in the following years as countries update their strategies to reflect increasing geopolitical tensions.

[6] It could be the case that “national security” occurs less than expected in national AI strategies as national security efforts may have their own national strategies of which AI’s use is a part of it. For example, the Government of Canada’s (2026) *Defence Industrial Strategy* includes AI use, development, and deployed throughout.

*Figure 4: National AI Strategy Policy-Goals' Convergence and Divergence*

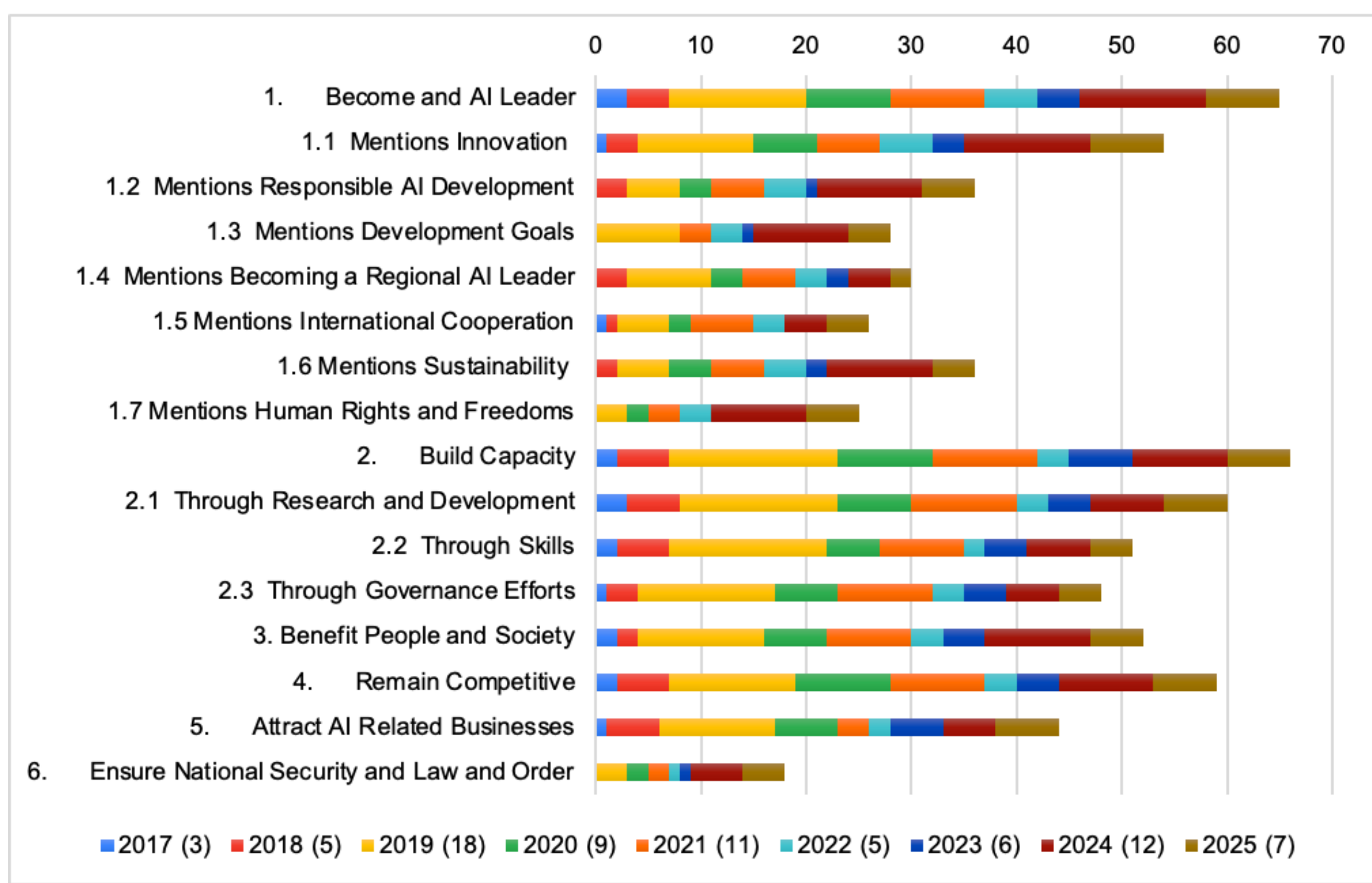

Figure 5 illustrates how national AI strategies adopt various policy-approaches over time, providing insight into the degree of horizontal convergence across countries. The most striking trend is the near-dataset-complete adoption of "Support National AI Research," which appears in 71/74 strategies. This broad code, along with its specific-codes (research, collaboration, and knowledge transfer) has steadily increased, suggesting that countries are converging on research and innovation as the basis of their AI policy-approach. For instance, Egypt's 2025 strategy highlights a coordinated industry-university research platform, while Ireland's 2024 strategy emphasizes collaborative networks as foundations for breakthroughs. These examples show convergence in AI policy rhetoric and institutionalization of similar implementation strategies. Other widely shared approaches, such as "Public Sector AI Adoption" and "Automate Processes, Increase Efficiency and Transparency," also demonstrate convergence. Countries like Bangladesh (2024) and Israel (2024) both describe automation as a pathway to improving public services and freeing workers for higher-value tasks. These trends suggest growing horizontal convergence in how governments envision AI's role in state modernization.

At the same time, divergence remains. Less frequently adopted policy-approaches, like "Encourage Public Consultations" or "Focused on Legislative and Regulatory Efforts" appear in fewer than 30 strategies. This limited uptake indicates weaker convergence around participatory and regulatory pathways, suggesting that while countries align on research and efficiency-oriented strategies, they diverge in how they involve citizens or pursue regulatory frameworks.

*Figure 5: National AI Strategy Policy-Approaches' Convergence and Divergence*

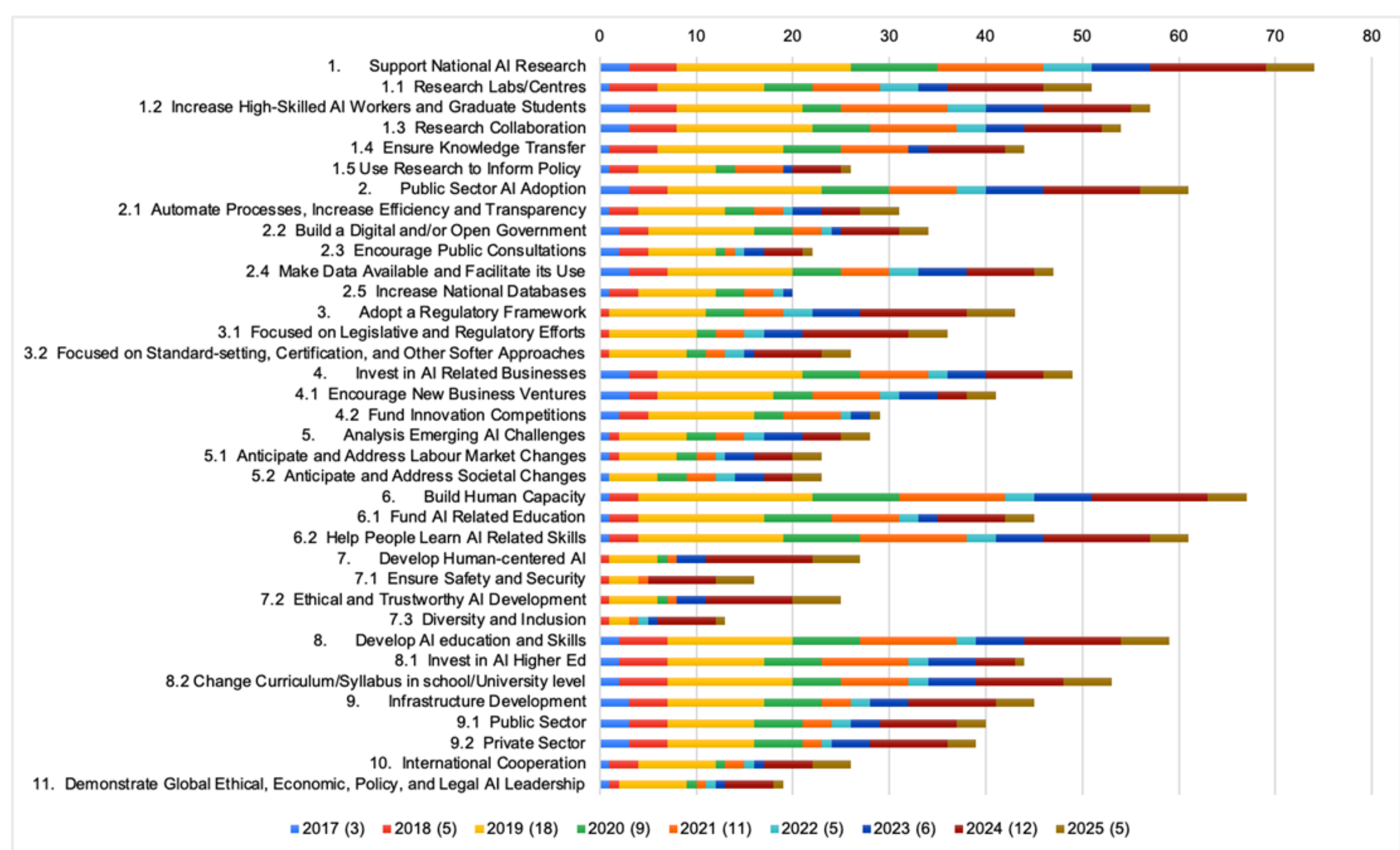

Figure 6 presents the distribution of policy-principles across national strategies, highlighting normative convergence. The principle of "Ethical AI Use" dominates, appearing in 68/74 strategies, demonstrating considerable horizontal convergence around the importance of ethical AI use. Specific codes, like privacy, transparency, and trust are consistently present, showing that countries are converging around a shared set of ethical guidelines. For instance, Nigeria's 2024 strategy emphasizes responsible data practices, while Spain's 2024 plan integrates privacy and cybersecurity into public trust frameworks. These examples, among others, reinforce that ethical commitments are becoming a policy-principle baseline. Beyond ethics, principles like "Explainability" and "Reliability" show notable increases after 2020, suggesting that policy convergence is extending beyond broad ethics to technical governance principles. The growth of these specific codes reflects in emerging horizontal convergence around transparency and accountability in AI systems.

However, divergence appears in principles tied to human values and inclusion, much like it did in policy-approaches above. For example, "Respects Human Dignity" and "Ensures Human Diversity and Inclusion" appears in fewer than 35 strategies and are unevenly distributed over time. Similarly, "Collective Debate" and "Use Media" are rarely mentioned. These absences reveal divergence in how far countries go in embedding societal inclusion and participatory governance into their AI strategies.

*Figure 6: National AI Strategy Policy-Principles' Convergence and Divergence*

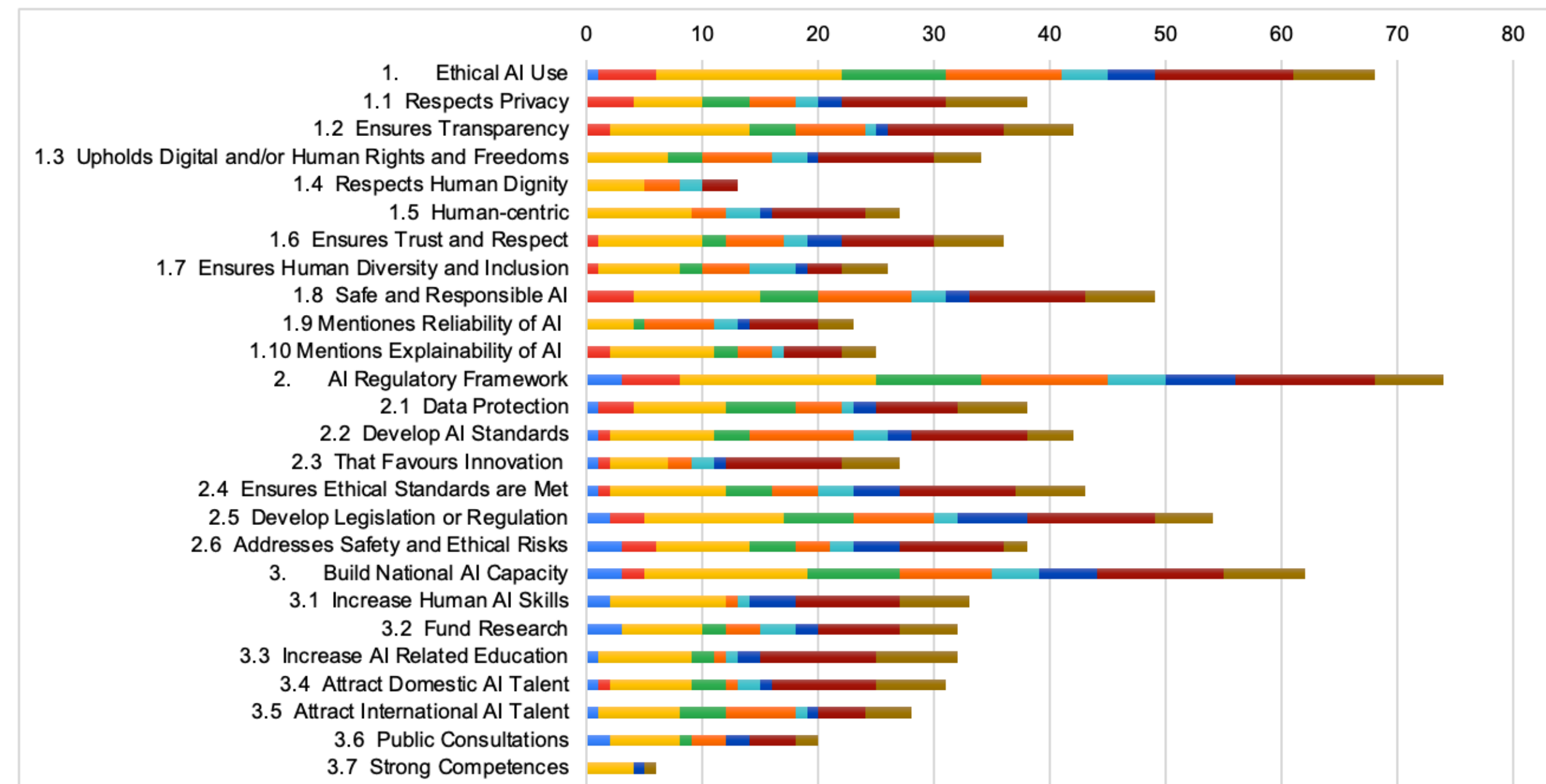

Turning to vertical patterns, we can look at the three regional strategies as they (mis)align with their member countries' strategies. The EU exhibits a mixed picture of vertical convergence. In terms of policy-goals (Figure 7), strong convergence is evident around economic ambition and capacity-building. Both the European Commission (EC) and member states emphasize becoming global AI leaders and building national capacity. For example, Ireland's 2024 strategy mirrors the EC's global leadership framing. However, again, divergence emerges in policy-goals related to human rights and national security, which member states mention more often than the EC. This indicates a partial but uneven convergence, where economic objectives are shared but social and security concerns diverge. Policy-approaches (Figure 8) reflect similar dynamics. EU Member states align closely with the EC on building human capacity and expanding AI-related skills, echoing Italy's 2024 focus on bridging the labour market skills gap. Yet, divergence appears in areas like diversity and inclusion, which the EC prioritizes but many states underemphasize. Conversely, some Member states innovate beyond the EC by advancing AI entrepreneurship and specific policy-approaches not found at the regional level. Thus, while vertical convergence is strong in skills and capacity-building, it is weaker in human-centric approaches. Policy-principles (Figure 9) show the greatest vertical convergence as the EC and Member states emphasize regulatory frameworks and ethical governance. For instance, France's strategy converges with the EC in calling for discrimination impact assessments and clear regulatory guidance. Still, divergence exists in policy-principles around diversity and inclusion, which remain less prominent at the national level. Overall, the EU demonstrates strong vertical convergence on regulatory and economic goals but divergence around human-centric values.

*Figure 7: Policy-Goals (EU national vs. regional)*

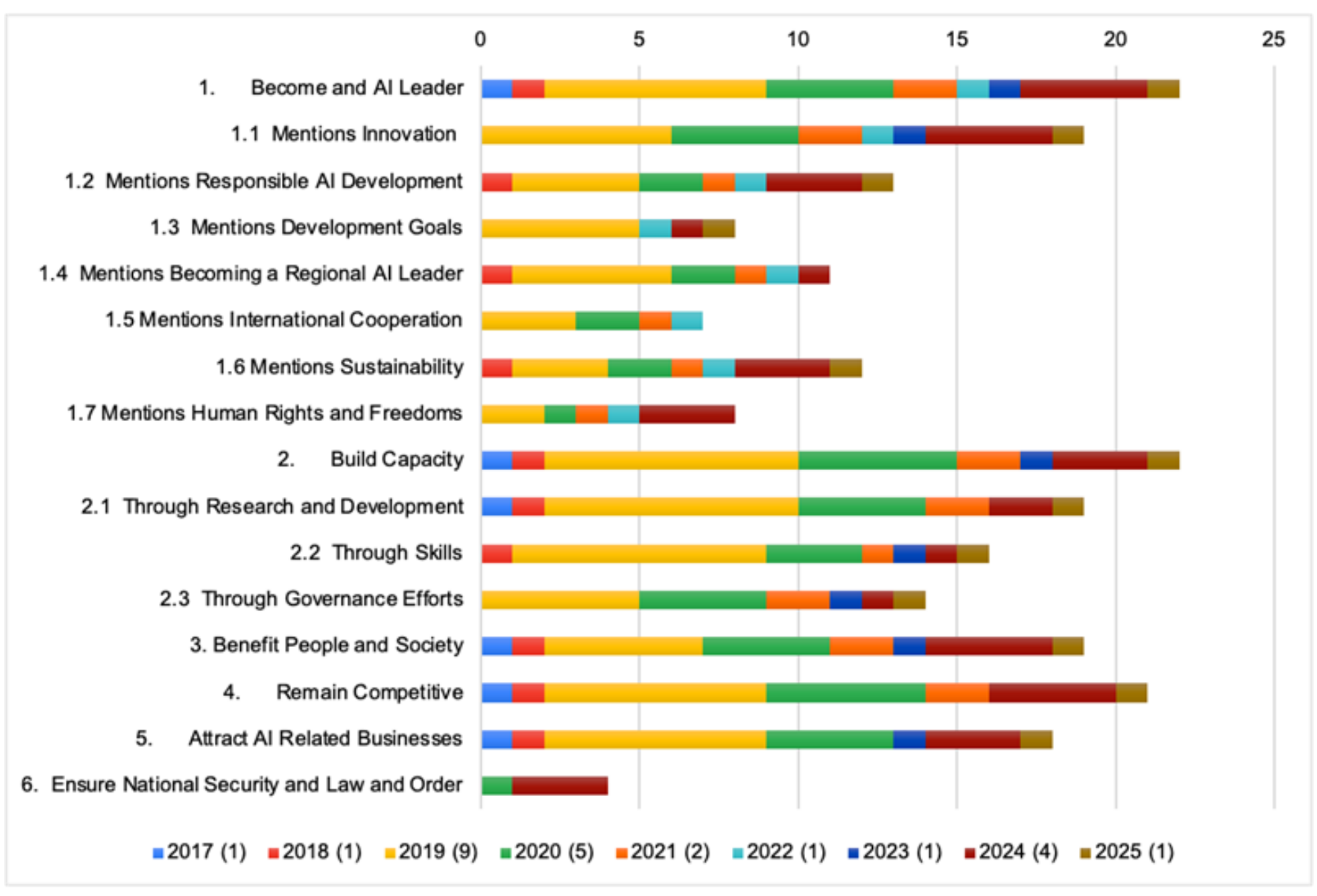

*Figure 8: Policy-Approaches (EU national vs. regional)*

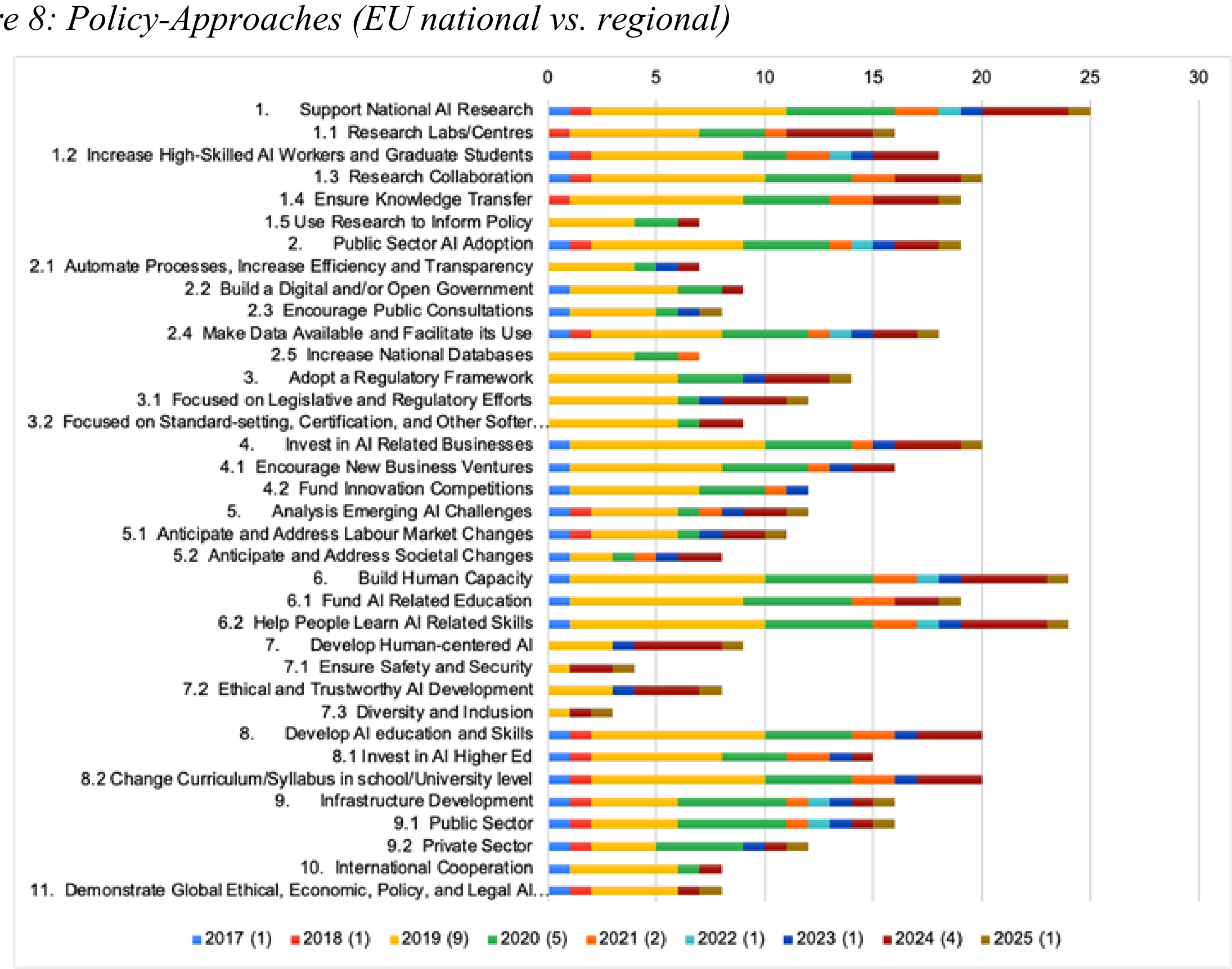

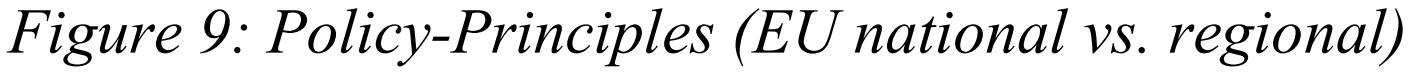

*Figure 9: Policy-Principles (EU national vs. regional)*

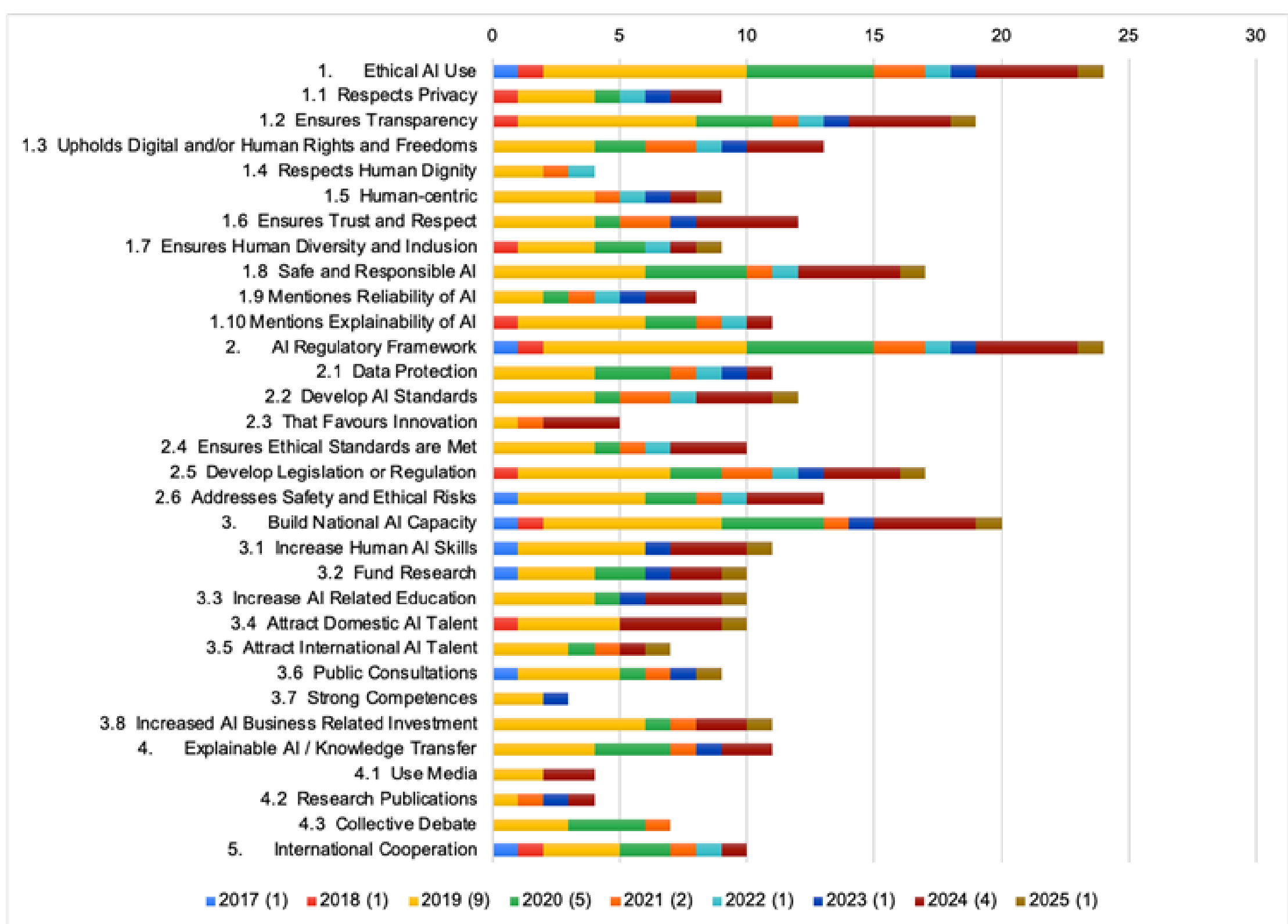

The AU shows strong vertical convergence across policy-goals, -approaches, and -principles. Most AU-endorsed policy-goals, like stimulating growth and solving societal challenges are echoed by its Member states (Figures 10-12). For example, Kenya's 2025 strategy includes ambitions to become Africa's leading AI hub, directly reflecting the AU's continental priorities. Minimal divergences were found, with just two national goals absent from the AU framework. Policy-approaches (Figure 11) reinforce this pattern. National to regional strategies converge on education, skills, and research as central pillars. For example, Egypt's strategy stresses supplying local labs with innovative technologies, which aligns with AU priorities. Only a handful of distinct national policy-approaches diverge from the AU's regional strategy, indicating strong convergence. Policy-principles (Figure 12) also strongly convergence in their emphasis of ethical use and data protection, like Nigeria's 2024 strategy exemplifying this shared commitment to transparency and accountability. Divergence appears only where some Member states introduce forward-looking principles like attracting international AI talent, which the AU has not yet codified. Overall, the AU region demonstrates the highest vertical convergence among the three regional-national strategies.

*Figure 10: Policy-Goals (AU national vs. regional)*

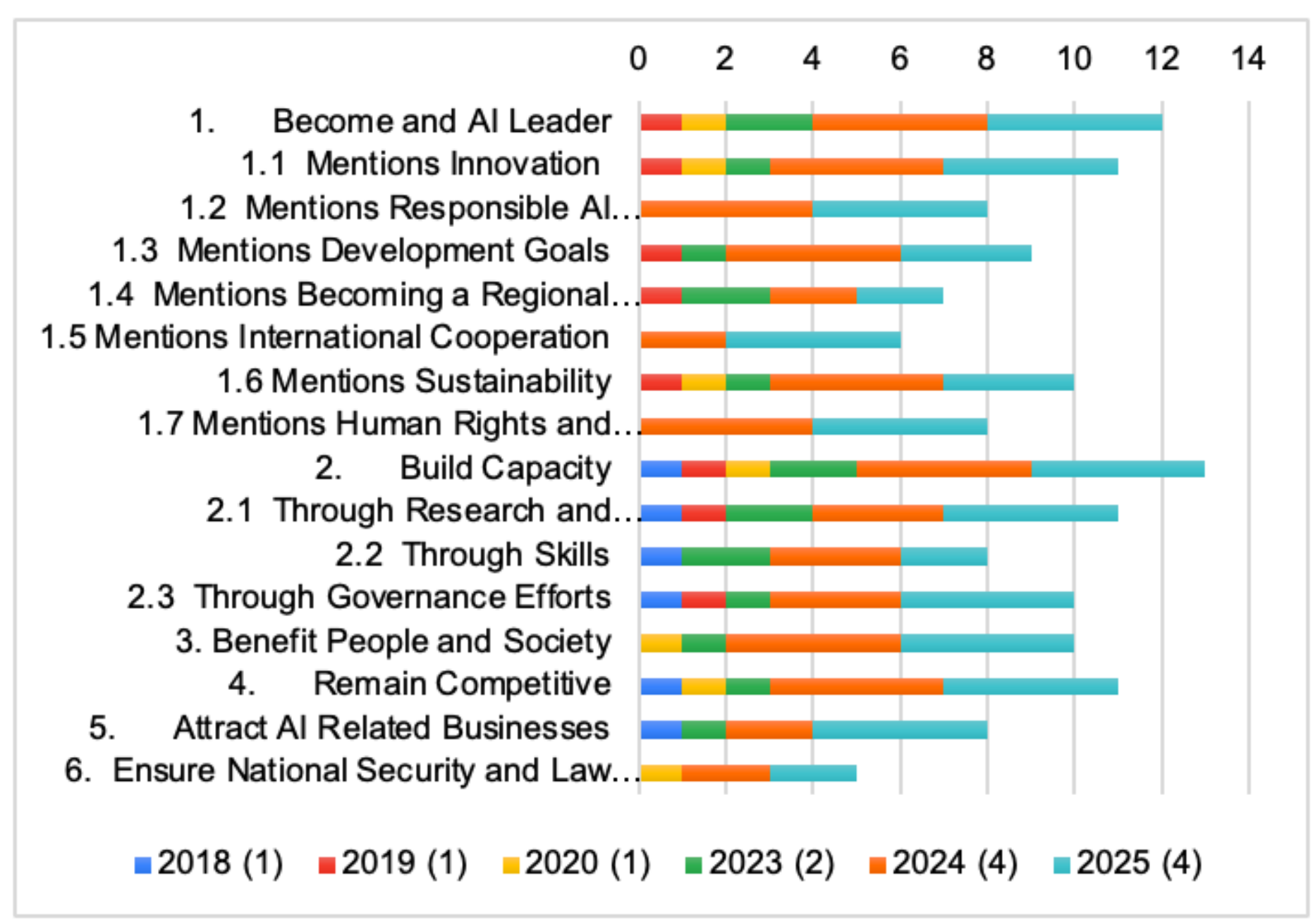

*Figure 11: Policy-Approaches (AU national vs. regional)*

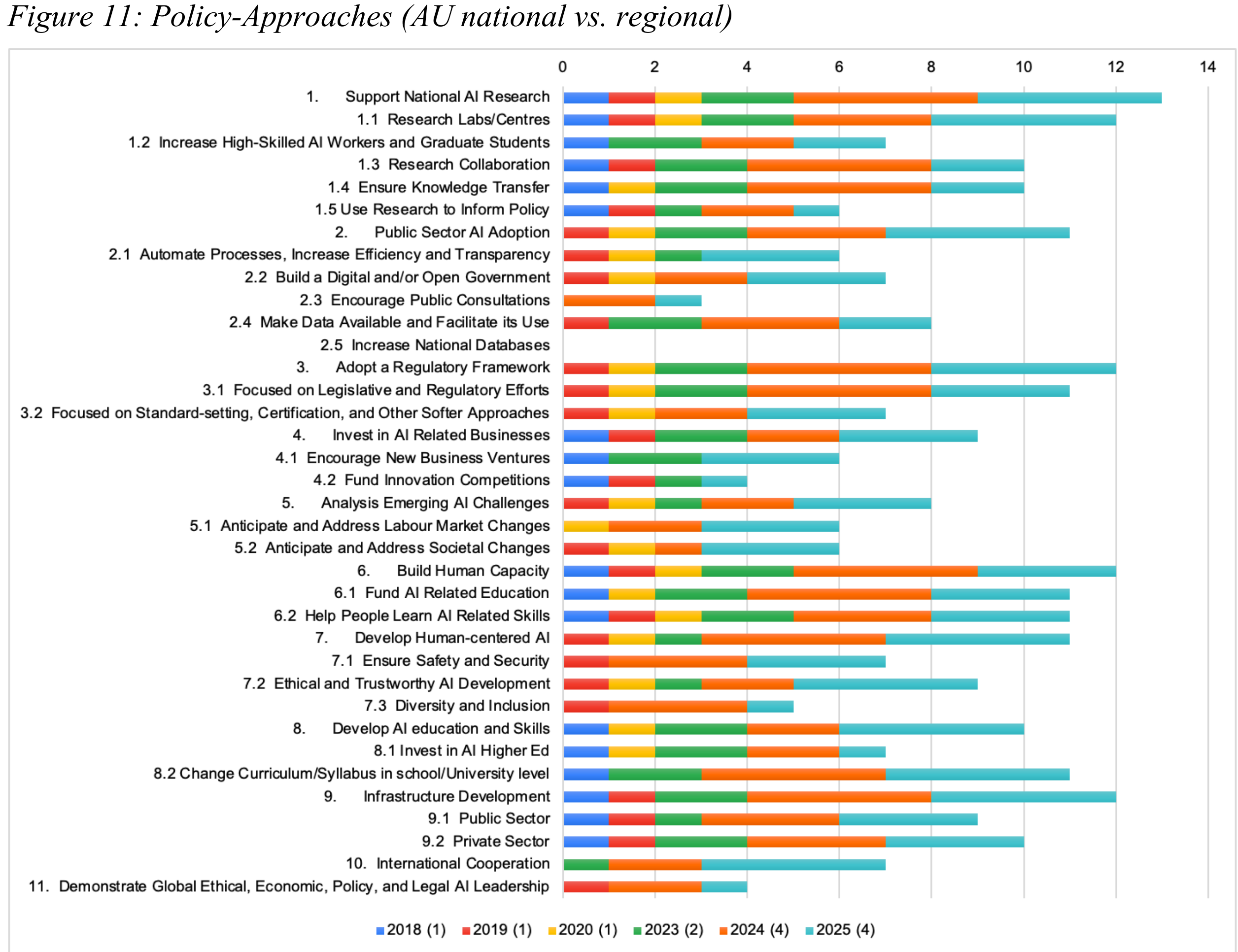

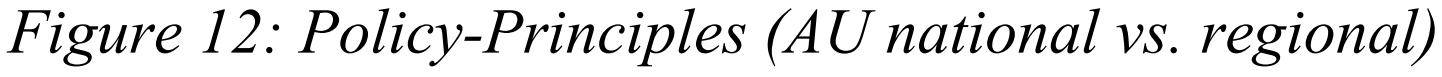

*Figure 12: Policy-Principles (AU national vs. regional)*

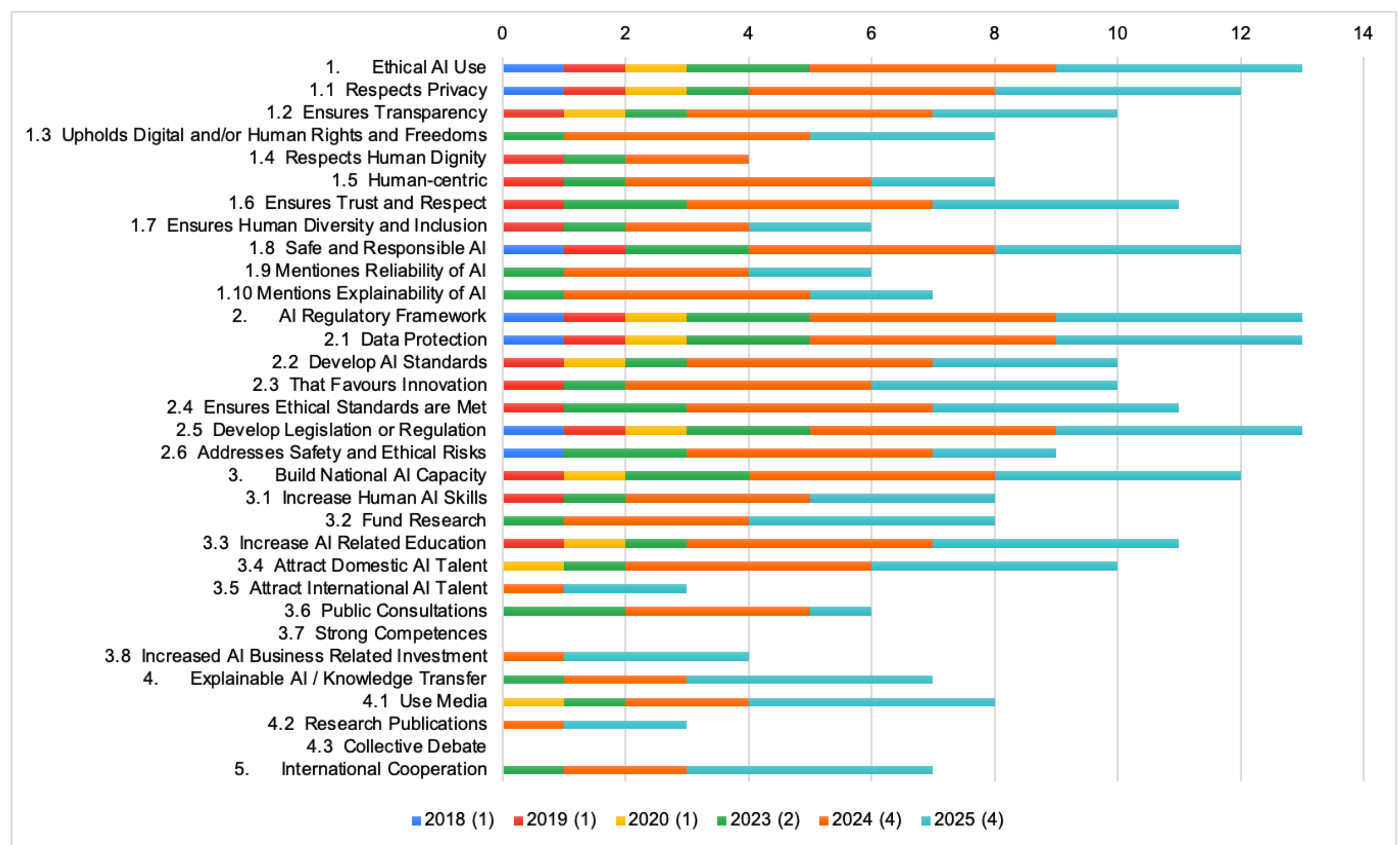

The NBR shows mixed vertical convergence (Figures 13-15). Policy-goals (Figure 13) reveal five shared priorities between regional and national levels. While both emphasize ethics and democracy, as seen in Denmark's 2019 principle of embedding ethics in AI, many national goals are absent at the regional level, producing mixed convergence. Policy-approaches (Figure 14) are similarly mixed. Only ten broad and specific policy-approaches are shared between the regional and various national AI strategies, and while some NBR states emphasize inclusive governance, others focus on innovation and entrepreneurship, which are not consistently reflected in the NBR AI strategy. This mix suggests a lack of harmonized vision compared to the EU and AU cases. Policy -principles (Figure 15) show modest convergence around transparency and ethics, but divergence elsewhere. National strategies often expand on principles not endorsed at the regional level, like talent attraction and research publication. As a result, vertical convergence in the NBR is mixed, with convergence in core ethical concerns but divergence in broader priorities.

*Figure 13: Policy-Goals (Nordic-Baltic national vs. regional)*

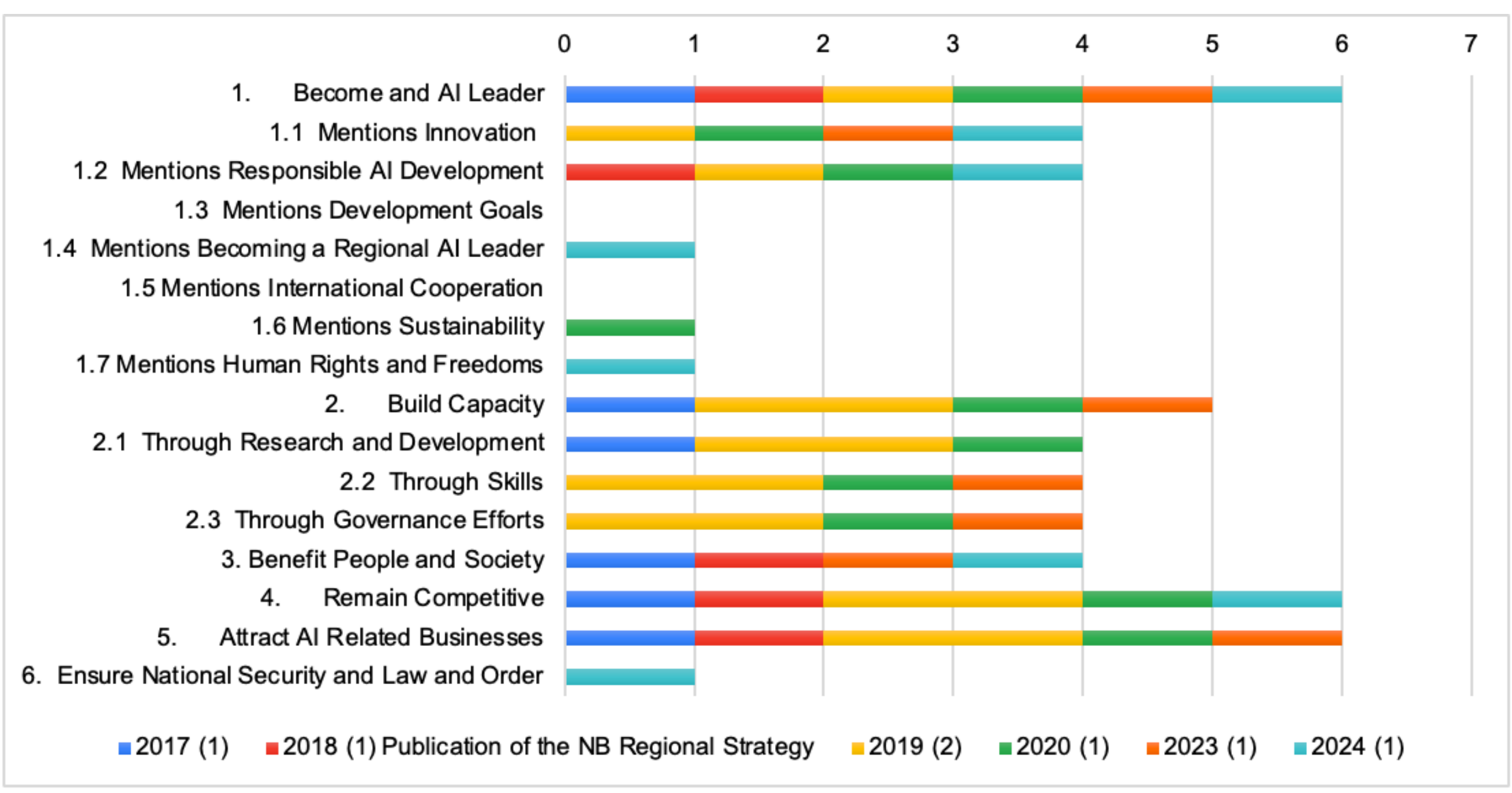

*Figure 14: Policy-Approaches (Nordic-Baltic national vs. regional)*

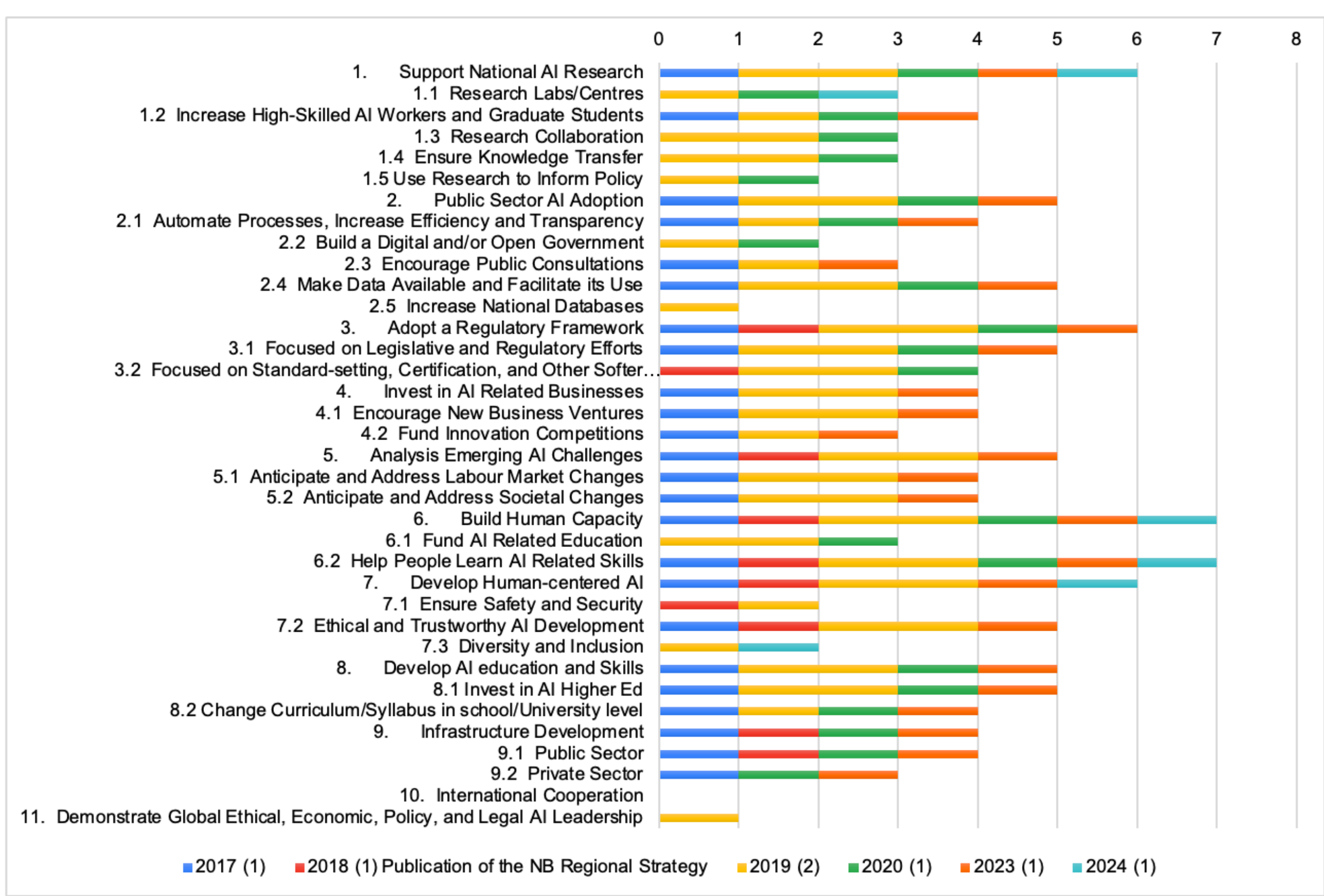

*Figure 15: Policy-Principles (Nordic-Baltic national vs. regional)*

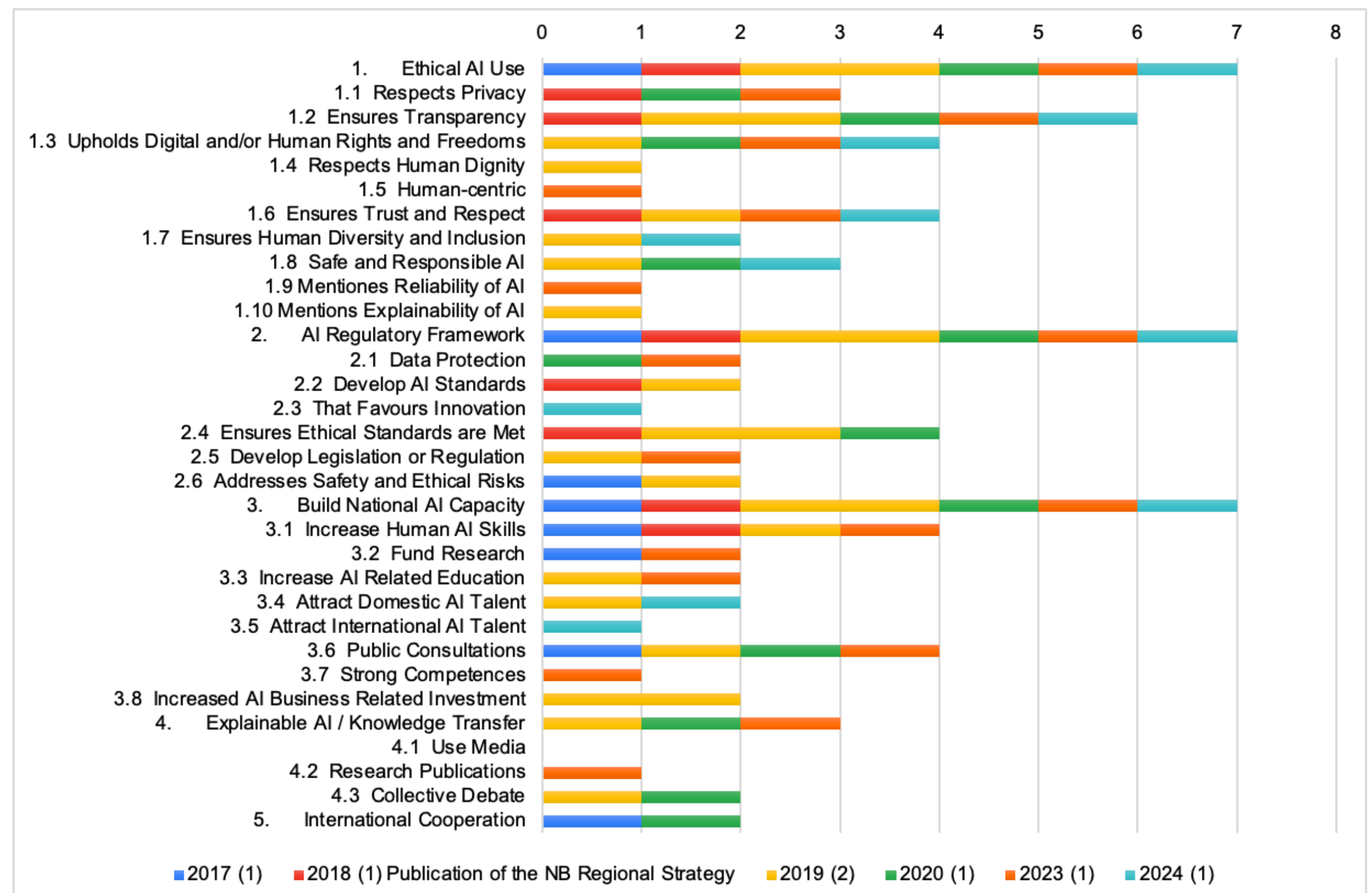

## Discussion

### *(a) Policy Design Element Variation*

This paper's results reveal considerable variation in the policy design element choices embedded across the 74 national AI strategies analyzed, with patterns of adoption that differ considerably across the three element categories – policy-goals, -approaches, and -principles – and across the broad and specific coding distinctions within each. Variation in policy-goals centres on a clear bifurcation between economic and social-related goals. Economic-related goals, like "Become an AI Leader" and "Build Capacity," dominate the landscape and remain consistent all years of strategy publication. These goals are typically elaborated through specific codes like "Through Research and Development," "Mentions Innovation," and "Through Skills," pointing to a broadly shared understanding of national AI policy as way to generate national productivity and competitive advantage. By contrast, human-rights and national security policy-goals appear in far fewer strategies (23 and 15 respectively) and do so unevenly across time and context. This variation maps onto the distinction Schneider and Sidney (2009) draw between policies that serve broad public interests and those that privilege specific target populations or political coalitions; the underrepresentation of rights-based goals suggests that while most states instrumentalize AI toward economic ends, the substantive integration of human rights as a policy-goal is still nascent.

Variation in policy-approaches is similarly structured. The near-complete adoption of "Support National AI Research" (71/74 strategies) and related specific codes for research collaboration and knowledge transfer suggests that the research-and-innovation pathway has functionally become the default policy-approach for national AI strategy design. However, substantial divergence emerges around more politically demanding approaches. For instance, "Encourage Public Consultations" and "Focused on Legislative and Regulatory Efforts" appear in fewer than 30 strategies, indicating that participatory and regulatory approaches diverge. This divergence is consequential, as Howlett & Rayner (2006) demonstrated precisely this dynamic in natural resource governance, where states converged on high-level rhetorical goals but diverged completely on the implementation approaches used to achieve them, a phenomenon the results here replicate in the AI policy context.

Variation in policy-principles is the most convergent finding whereby "Ethical AI Use" appears in 68/74 strategies and is accompanied by consistent sub-codes around privacy, transparency, and trust, suggesting a near-baseline consensus on the importance of ethical governance in AI policy design at the national-level. At the same time, principles related to "Ensures Human Diversity and Inclusion" and "Respects Human Dignity" appear in fewer than 35 strategies and are distributed unevenly over time, revealing that the normative commitment to ethical AI, while broad, is substantively shallow in a large proportion of strategies. This distinction between broad convergence on ethical rhetoric and divergence on human-centric principles is precisely the divergence-within-convergence dynamic that Howlett & Rayner (2006) identified, where states adopt similar high-level governance language without producing equivalent substantive policy outputs. These principles-level variations also raise concerns about policy layering and drift over time, since a broad consensus on ethics that is not reinforced by specific human-centric commitments may produce increasingly divergent policy mixes as strategies are updated, a risk that Drezner (2005) and Howlett & Rayner (2006) both flag as a key challenge for policy designers working in fast-moving policy domains.

*(b) Horizontal and Vertical Policy Convergence and Divergence Implications*

The horizontal convergence findings carry considerable implications for the emerging AI policy landscape. The strong and accelerating convergence around economic competitiveness and research-oriented goals, particularly after 2020, is consistent with Drezner's (2001) observation that convergence driven by early-mover states – in this case Canada, China, and Finland – define the strategic language that subsequent strategies adopt, whether through lesson-drawing, regulatory competition, or voluntary adoption. Bennett (1991) similarly notes that convergence driven by emulation produces patterns where the direction of change is identifiable even if the causal mechanism varies by policy. The temporal trends in this dataset, with convergence clearly accelerating post-2020, support the view that the international diffusion of AI strategies is a structured process of cumulative adoption. Conversely, the horizontal divergence around human-centric goals, participatory approaches, and inclusion-oriented policy-principles implies that the emerging AI policy paradigm, while increasingly convergent in its economic framing, remains divergent in its social commitments. For policy designers, this horizontal divergence is a risk factor as jurisdictions operating at the same administrative level cannot converge on human-centric principles.

The vertical convergence and divergence results across the three regions reveal how regional variations may contribute to the translation of supranational priorities into national policy design elements. For instance, EU Member state strategies converge strongly with the EC on economic objectives and regulatory principles, converging with the EC's consistent emphasis on building a competitive, trustworthy, and regulatory-compliant AI ecosystem. This pattern of strong vertical convergence on regulatory frameworks is consistent with what Zhu et al. (2025)

identify as institutional channelling, whereby the hierarchical authority of a supranational body, like the EU, shapes Member state policy design choices through formal coordination mechanisms and the broad regulatory shadow of instruments like the EU *AI Act*. However, EU Member state diverges on diversity and inclusion principles, a domain the EC prioritizes but most national strategies do not, illustrates a tension between the EC's supranational human-centric agenda and the more economically focused domestic preferences of member states, echoing the divergence-within-convergence dynamic discussed above.

The AU's high vertical convergence across all three policy design element categories presents a markedly different regional dynamic. The near-complete convergence of national strategies with the AU's continental priorities around growth, societal benefit, ethical use, and data protection indicates that regional strategy has been an effective template for national AI strategy development across AU member states. This high convergence may reflect a different mechanism than the EU case where rather than institutional hierarchy or regulatory pressure, the AU pattern may be driven by the limited independent AI policy capacity of individual member states, which leads them to draw heavily on the regional framework as a foundational document. Bennett's (1991) lesson-drawing model is particularly applicable here, as the AU's strategy functions less as a binding mandate and more as a shared policy blueprint from which member states can systematically draw from, a process that naturally produces high convergence in policy design element choices. The few points of divergence that do emerge, like principles around international talent attraction, suggest that more advanced member states are beginning to supplement the regional baseline with context-specific elements, which may produce gradual divergence over time as national capacity grows.

The NBR's mixed vertical convergence stands in contrast to both the EU and AU patterns and raises important questions about the conditions under which regional strategies produce convergent national policies. The NBR's limited shared policy-approaches and the tendency for Member states to develop nationally specific priorities outside the regional strategy's scope suggests that the NBR strategy functions more as a declaration of shared values rather than an operational governance document. Holzinger & Knill (2005) caution that convergence at the level of broad principles does not guarantee convergence at the level of implementation approaches, and the NBR case exemplifies this, where shared ethical and democratic commitments among NBR member states do not translate into convergent policy-goals or -approaches. For regional policy designers, this finding underscores the risk of designing regional AI strategies at too high a level of abstraction, as strategies that lack operational specificity may produce minimal convergence while leaving member states to develop substantively divergent national approaches that could conflict in areas like data governance, regulatory standards, or AI procurement.

*(c) Utility of Results for Policy Designers: Future National and Regional AI Strategy Efforts*

The utility of these findings for policy designers is fourfold. First, systematic policy convergence analysis enables policy designers to engage in what Bennett (1991) called "lesson-drawing" – the deliberate and structured analysis of first-mover states' policies to design better and more informed domestic policies. When policy designers can observe *what* policy design elements other countries have adopted, *when* they adopted them, and whether those adoptions have become widespread compared to other countries' efforts, they gain access to a richer evidence base for evaluating their own policy design choices. For instance, a policy designer in a late-adopting country can assess which policy-approach codes have converged, like "Support National AI Research" and which ones diverged to calibrate their strategy accordingly.

Second, understanding *where* convergence is and is not occurring allows policy designers to identify emerging norms that may harden into assumed policy design standards. As Drezner (2001) argued, great powers play a disproportionate role in determining the direction of policy convergence, when the US and EU act in concert, the resulting policy framework tends to propagate across other jurisdictions through a combination of market pressure, institutional mimicry, and deliberate harmonization. For AI policy designers outside these great power blocs, recognizing which convergence are being driven by powerful actors and which are the product of more genuinely distributed consensus is critical for assessing the degree of flexibility available in ongoing AI policy design choices.

Third, tracking temporal convergence and divergence across all three policy design element types helps policy designers avoid the trap of policy layering, drift, and incoherence. Howlett (2014) argued that without attention to the sequencing and temporal evolution of policy mixes, designers risk accumulating redundant, contradictory, or counter-productive policy instruments over time as each successive update adds new elements without pruning old ones. By mapping which design elements have converged and diverged, and which display the divergence within convergence pattern identified by Howlett & Rayner (2006), designers can proactively identify tensions within their existing strategies and take corrective action before those tensions crystallize into governance failures. The persistent divergence observed in human-centric principles across national AI strategies illustrates precisely this risk whereby nations may adopt similar language without making equivalent substantive commitments, creating a form of ethics washing that can erode public trust and generate policy conflict as implementation proceeds. Understanding this divergence pattern alerts policy designers to the need for more explicit and operationally specific commitments in these areas.

Fourth, vertical convergence analysis gives policy designers working within regional governance frameworks, whether EU, AU, or NBR Member states a precise understanding of the degree to which their national strategies converge or diverge from the regional strategy. This understanding is particularly consequential in contexts like the EU, where the EU *AI Act* is set to define regulatory implementation across Europe and influence partners well beyond their Member states. In aggregate, a comprehensive, temporally sensitive, and spatially differentiated analysis of policy design element convergence and divergence across national and regional AI strategies provides policy designers with an useful knowledge base that clarifies what is happening in the policy area, what is regionally specific, what is converging and diverging, and what is still being worked out, enabling a more deliberate and strategically informed AI policy design choices.

## Conclusion

The proliferation of national and regional AI strategies since 2017 has created opportunity and complexity for policy designers tasked with shaping their countries national AI policy efforts. This paper addressed that complexity by constructing a comprehensive dataset of 74 national and 3 regional AI strategies drawn from a global scan of all 205 UN member and non-member states, applying a latent-inductive coding approach organized around the three functional policy design elements: policy-goals, -approaches, and -principles, and charting the temporal shifts of those coded elements to evaluate horizontal and vertical of policy convergence and divergence. The result a descriptive inventory of what AI strategies contain, and a structured account of how these strategies are evolving in relation to one another and what that evolution signals for future AI policy design choices. For policy designers, the practical value of this kind of systematic, comparative, and temporally sensitive analysis is substantial. Rather than approaching AI strategy development in isolation or relying on ad-hoc comparisons with a handful of prominent national strategies, designers can draw on the full breadth of what over seven years of published strategies reveal, which policy design elements have achieved near-complete uptake or remain contested or unevenly distributed, and which are accelerating toward convergence in ways that suggest emerging international norms around policy design element choices or mixes. For instance, the near-complete adoption of research support as a policy-approach and ethical AI use as a policy-principle marks these elements as de-facto baseline expectations that any credible new or updated strategy would be expected to address. Simultaneously, the persistent underrepresentation of human rights, participatory governance, diversity and inclusion, and human dignity across national and regional AI strategies exposes a structural gap between the commitments embedded in AI strategies and the substantive policy commitments that would operationalize those values. Recognizing this divergence-within-convergence dynamic, whereby nations converge on high-level language without producing equivalent human-centric policy outputs, gives policy designers a precise diagnostic for identifying where ethics washing risks eroding public trust and for designing more substantively grounded efforts in future strategy updates. Similarly, the accelerating pace of convergence, as observed after 2020, suggests that the window for independent problem-solving is narrowing; late-adopting states that have not yet locked in their strategic priorities face increasing structural pressure to converge with frameworks shaped by early-mover countries.

Notwithstanding these contributions, this paper carries three limitations that warrant careful consideration when applying its findings. First, its temporal and linguistic scope. At the time of data collection, several national AI strategies (30) were still in progress, with an additional five called for but not yet published. As these strategies are released, the broader trends and findings reported in this paper may evolve. Furthermore, due to the language accessibility constraints, 11 non-English strategies were excluded from this analysis. This exclusion introduces the possibility of overlooking unique policy-goals, -approaches, or -principles in non-English contexts, thereby limiting the generalizability of the coding outcomes. Second, measuring policy convergence is difficult and does not have a set methodology; other studies have used case studies, quantitative policy analysis, among others to determine varying policy convergence levels across policy domains. While this research does its best to measure policy convergence across a policy domain using an inductive set of codes, it is still an imperfect policy convergence measure. Third, plotting broad and specific policy design element codes' total frequency against publication year can misrepresent findings. The 74 included national AI strategies were most often published in 2019, making 2019's results appear more frequently than other years (2017 = 3, 2018 = 5, 2019 = 18, 2020 = 9, 2021 = 11, 2022 = 5, 2023 = 6, 2024 = 12, 2025 = 5). Reducing each broad and specific theme frequency to a percentage of total national AI strategies published within each year would normally address this limitation. However, since so few strategies exist across so few years, this limitation addressing method is insufficient as it may disproportionately represent some policy design element convergence or divergence, such as if all three strategies in 2017 included or excluded a given policy design element code – making it read as 0 or 100%.

Several productive avenues for future research emerge from these findings. First, as the cohort of in-progress strategies is released and existing strategies are updated, a longitudinal extension of this dataset would enable researchers to test whether the accelerating convergence trends observed through 2025 continue shift, particularly as geopolitical and economic tensions increasingly shape AI policy discourse. Second, the non-English strategy corpus that was excluded here represents a meaningful gap; future research incorporating translated versions of these documents would produce a more globally representative picture of AI strategy design, potentially surfacing convergence and divergence patterns that are currently missing. Third, while this study identifies the *what* and *when* of convergence, the causal mechanisms remain underexplored in the national and regional AI strategy context. Fourth, expanding vertical convergence analysis to sub-national or federal-provincial contexts, such as comparing Canadian provincial AI governance efforts with the federal strategy, would extend this paper's analytical reach into governance hierarchies beyond the supranational level. Finally, as new and updated AI strategies place growing emphasis on compute access, sovereign data infrastructure, and AI safety frameworks, future coding efforts could add these emerging design elements to the codebook to capture the evolving priorities in the AI policy area.

## Appendices

*Appendix A: Status of National and Regional AI Strategies*

| Status/Scope | Country | Region |
|---|---|---|
| **Yes (language if non-English)** | Argentina (Spanish); Australia; Austria (German); Azerbaijan (Azerbaijani); Bangladesh; Belgium; Benin; Brazil; Bulgaria (Bulgarian); Canada; Chile (Spanish); China; Colombia (Spanish); Costa Rica (Spanish); Côte d'Ivoire (French); Cyprus (Greek); Czechia (Czech); Denmark; Dominican Republic (Spanish); Egypt; Estonia (English); Ethiopia; Finland; France; Germany; Greece (Greek); Hungary; Iceland (Icelandic); India; Indonesia (Indonesian); Ireland; Israel; Italy; Jamaica; Japan; Jordan; Kenya; Kuwait; Lebanon; Lesotho; Lithuania; Luxembourg; Malaysia; Malta; Mauritania; Mauritius; Mexico; Nepal (Nepali); Netherlands; Nigeria; Norway; Oman; Pakistan; Peru; Philippines; Poland; Portugal; Qatar; Romania (Romanian); Russia; Rwanda; Saudi Arabia; Serbia; Singapore; Slovakia; Slovenia (Slovenian); South Africa; South Korea; Spain; Sri Lanka; Sweden; Switzerland (German); Tajikistan; Taiwan (Mandarin Chinese); Thailand; Turkey; Ukraine; United Arab Emirates; United Kingdom; United States of America; Uruguay (Spanish); Uzbekistan; Zambia | African Union; European Union; Nordic-Baltic Regions |
| **Yes, but inaccessible** | Algeria; Bahrain; Croatia; Ghana; Iran; Latvia; Namibia; Palestine State; Senegal; Sierra Leone | |
| **In-Progress** | Albania; Andorra; Antigua and Barbuda; Armenia; Bahamas; Barbados; Belarus; Bhutan; Cambodia; Cameroon; Cuba; Djibouti; Eswatini; Georgia; Honduras; Iraq; Kazakhstan; Kyrgyzstan; Laos; Mongolia; Montenegro; Morocco; Mozambique; Myanmar; New Zealand; North Macedonia; Papua New Guinea; Tanzania; Togo; Tunisia; Vietnam; Zimbabwe | |
| **Called For** | Angola; Gabon; Somalia; Trinidad and Tobago; Uganda | |
| **No** | Abkhazia; Afghanistan; Belize; Bolivia; Bosnia and Herzegovina; Botswana; Brunei; Burkina Faso, Burundi, Cabo Verde; Central African Republic; Chad; Comoros; Congo-Brazzaville; Cook Islands; Democratic Republic of the Congo; Dominica; Ecuador; El Salvador; Equatorial Guinea; Eritrea; Fiji; Gambia; Grenada; Guatemala; Guinea; Guinea-Bissau; Guyana; Haiti; Holy See; Kiribati; Kosovo; Liberia; Libya; Liechtenstein; Madagascar; Malawi; Maldives; Mali; Marshall Islands; Micronesia; Moldova; Monaco; Nauru; Nicaragua; Niger; Niue; Northern Cyprus; North Korea; Palau; Panama; Paraguay; Saint Kitts and Nevis; Sahrawi Arab Democratic Republic; Saint Lucia; Saint Vincent and the Grenadines; Samoa; San Marino; Sao Tome and Principe; Seychelles; Solomon Islands; Somaliland; South Ossetia; South Sudan; Sudan; Suriname; Syria; Timor-Leste; Tonga; Transnistria; Turkmenistan; Tuvalu; Vanuatu; Venezuela; Yemen | |

*Appendix B: Identified Board and Specific Policy Design Element in AI Strategies*

| Policy Design Element | Broad / Specific | Codes |
|---|---|---|
| Goal | Broad | 1. Become an AI Leader |
| | Specific | 1.1 Mentions Innovation |
| | | 1.2 Mentions Responsible AI Development |
| | | 1.3 Mentions Development Goals |
| | | 1.4 Mentions Becoming a Regional AI Leader |
| | | 1.5 Mentions International Cooperation |
| | | 1.6 Mentions Sustainability |
| | | 1.7 Mentions Human Rights and Freedoms |
| | Broad | 2. Build Capacity |
| | Specific | 2.1 Through Research and Development |
| | | 2.2 Through Skills |
| | | 2.3 Through Governance Efforts |
| | Broad | 3. Benefit People and Society |
| | Broad | 4. Remain Competitive |
| | Broad | 5. Attract AI Related Businesses |
| | Broad | 6. Ensure National Security and Law and Order |
| Approach | Broad | 1. Support National AI Research |
| | Specific | 1.1 Research Labs/Centers |
| | | 1.2 Increase High-Skilled AI Workers and Graduate Students |
| | | 1.3 Research Collaboration |
| | | 1.4 Ensure Knowledge Transfer |
| | | 1.5 Use Research to Inform Policy |
| | Broad | 2. Public Sector AI Adoption |
| | Specific | 2.1 Automate Processes, Increase Efficiency and Transparency |
| | | 2.2 Build a Digital and/or Open Government |
| | | 2.3 Encourage Public Consultations |
| | | 2.4 Make Data Available and Facilitate its Use |
| | | 2.5 Increase National Databases |
| | Broad | 3. Adopt a Regulatory Framework |
| | Specific | 3.1 Focused on Legislative and Regulatory Efforts |
| | | 3.2 Focused on Standard-setting, Certification, and Other Softer Approaches |
| | Broad | 4. Invest in AI Related Businesses |
| | Specific | 4.1 Encourage New Business Ventures |
| | | 4.2 Fund Innovation Competitions |
| | Broad | 5. Analysis Emerging AI Challenges |
| | Specific | 5.1 Anticipate and Address Labour Market Changes |
| | | 5.2 Anticipate and Address Societal Changes |
| | Broad | 6. Build Human Capacity |
| | Specific | 6.1 Fund AI Related Education |
| | | 6.2 Help People Learn AI Related Skills |
| | Broad | 7. Develop Human-centered AI |
| | Specific | 7.1 Ensure Safety and Security |
| | | 7.2 Ethical and Trustworthy AI Development |
| | | 7.3 Diversity and Inclusion |
| | Broad | 8. Develop AI education and Skills |
| | Specific | 8.1 Invest in AI Higher Ed |
| | | 8.2 Change Curriculum/Syllabus in school/University level |
| | Broad | 9. Infrastructure Development |
| | Specific | 9.1 Public Sector |
| | | 9.2 Private Sector |
| | Broad | 10. International Cooperation |
| | Broad | 11. Demonstrate Global Ethical, Economic, Policy, and Legal AI Leadership |
| Principle | Broad | 1. Ethical AI Use |
| | Specific | 1.1 Respects Privacy |
| | | 1.2 Ensures Transparency |
| | | 1.3 Upholds Digital and/or Human Rights and Freedoms |
| | | 1.4 Respects Human Dignity |
| | | 1.5 Human-centric |
| | | 1.6 Ensures Trust and Respect |
| | | 1.7 Ensures Human Diversity and Inclusion |
| | | 1.8 Safe and Responsible AI |
| | | 1.9 Mentions Reliability of AI |
| | | 1.10 Mentions Explainability of AI |
| | Broad | 2. AI Regulatory Framework |
| | Specific | 2.1 Data Protection |
| | | 2.2 Develop AI Standards |
| | | 2.3 That Favours Innovation |
| | | 2.4 Ensures Ethical Standards are Met |
| | | 2.5 Develop Legislation or Regulation |
| | | 2.6 Addresses Safety and Ethical Risks |
| | Broad | 3. Build National AI Capacity |
| | Specific | 3.1 Increase Human AI Skills |
| | | 3.2 Fund Research |
| | | 3.3 Increase AI Related Education |
| | | 3.4 Attract Domestic AI Talent |
| | | 3.5 Attract International AI Talent |
| | | 3.6 Public Consultations |
| | | 3.7 Strong Competences |
| | | 3.8 Increased AI Business Related Investment |
| | Broad | 4. Explainable AI / Knowledge Transfer |
| | Specific | 4.1 Use Media |
| | | 4.2 Research Publications |
| | | 4.3 Collective Debate |
| | Broad | 5. International Cooperation |